\documentclass[12pt]{article}
\usepackage{graphicx}
\advance\oddsidemargin by -\textwidth
\advance\oddsidemargin by -1in
\evensidemargin=\oddsidemargin
\def\fsl#1{\setbox0=\hbox{$#1$}           
   \dimen0=\wd0                                 
   \setbox1=\hbox{/} \dimen1=\wd1               
   \ifdim\dimen0>\dimen1                        
      \rlap{\hbox to \dimen0{\hfil/\hfil}}      
      #1                                        
   \else                                        
      \rlap{\hbox to \dimen1{\hfil$#1$\hfil}}   
      /                                         
   \fi}                                         %
\usepackage{amsmath,amssymb}
\usepackage{array}
\newcommand{\VEV}[1]{\langle \overline{#1} {#1} \rangle}
\newcommand{\vev}[1]{\langle #1 \rangle}

\newcommand{\TeV}{\text{TeV}}
\newcommand{\GeV}{\text{GeV}}
\newcommand{\MeV}{\text{MeV}}
\newcommand{\crit}{\rm crit}

\newcommand{\asymm}{\rm asym}
\newcommand{\beq}{\begin{eqnarray}}
\newcommand{\eeq}{\end{eqnarray}}
\makeatletter
\def\@maketitle{\newpage
 \null
 {\normalsize \tt \begin{flushright} 
  \begin{tabular}[t]{l} \@date 
  \end{tabular}
 \end{flushright}}
 \begin{center}
 \vskip 2em
 {\LARGE \@title \par} \vskip 2.5em {\large \lineskip .5em
 \begin{tabular}[t]{c}\@author 
 \end{tabular}\par} 
 \end{center}
 \par
 \vskip 1.5em} 
\makeatother
\makeatletter
\title{{\Large\bf
     Anatomy of  Top-Mode Extended Technicolor Model
\footnote{A preliminary version is published in 
the Proceedings of the 2006 International Workshop: 
Origin of Mass and Strong Coupling Gauge Theories (SCGT 06), Nov. 21-24, 2006, 
Nagoya University, Nagoya, eds M. Harada, M. Tanabashi and K.Yamawaki 
(World Scientific Publishing Co., Singapore 2007)}}}
\author{
  {\large
    Hidenori {\sc Fukano}\thanks{
      {\tt fukano@eken.phys.nagoya-u.ac.jp}}\quad
    and \quad
    Koichi {\sc Yamawaki}\thanks{
      {\tt yamawaki@eken.phys.nagoya-u.ac.jp}}
  }\\[5mm]
  {\it Department of Physics, Nagoya University} \\ 
  {\it Nagoya 464-8602, Japan}
}
\date{\today}
\begin{document}
\maketitle
\begin{abstract}
We analyze two versions of the extended technicolor (ETC) 
incorporating the top quark condensate via the flavor-universal coloron type topcolor $SU(3)_1 \times SU(3)_2$: 
A straightforward top-mode ETC having quarks and techniquarks assigned to a single (strong) $SU(3)_1$, 
and a ``twisted model'' with techniquarks carrying the weak $SU(3)_2$ while quarks the strong $SU(3)_1$.  
The straightforward model has the same ETC structure as that of Appelquist et al. 
without topcolor which we first analyze to find  
that it yields only too small ETC-induced mass for the third generation. 
In contrast, our model having topcolor takes 
the form of a version of the topcolor-assisted technicolor (TC2) after ETC breakings, 
which triggers the top quark condensate giving rise to a realistic top mass.  
However, techniquarks have the strong topcolor $SU(3)_1$ 
in addition to the already strong walking/conformal technicolor, 
which triggers the techniquark condensate at scale much higher than the weak scale, a disaster.  
We then consider a ``twisted model'' of TC2, though not an explicit ETC.
We find  
a new feature that ``ETC''-induced quark mass is enhanced to 
the realistic value 
by the large anomalous dimension $\gamma_m \simeq 2$ of 
Nambu-Jona-Lasinio-type topcolor interactions. 
The result roughly reproduces the realistic quark masses. 
We further find  a novel effect of 
the above large anomalous dimension $\gamma_m \simeq 2$: 
The  top-pion mass has a universal upper bound,  $m_{\pi_t} < 70\,\GeV$,
in the generic TC2 model. 
\end{abstract}
\newpage

\section{Introduction}
\label{intro}

The Standard model (SM) has a mysterious part, the electroweak symmetry breaking (EWSB),
to give masses of quarks, leptons and W/Z-bosons. 
The EWSB via the elementary Higgs field in the SM has some problems. 
In order to solve these problems, we should build a scenario beyond the SM. 
One of candidates for such scenarios is Technicolor (TC)~\cite{Weinberg:1975gm}.
TC is an attractive idea for the EWSB without the elementary Higgs, based on an analogy 
to the QCD with technifermion condensate, 
instead of the quark condensate in QCD, responsible for the mass of  W/Z bosons.
In order to give mass to the SM fermions, 
we have to extend the TC into a larger picture which communicates
the technifermion to the SM fermion. 
A typical one is the extended TC (ETC) model~\cite{Dimopoulos:1979es} in which the TC group 
is embedded into a larger gauge group including the horizontal gauge group of three families of the SM fermion.%
\footnote{
Another possibility is a composite model 
where both the SM fermions and the technifermions are composite on equal footing, 
in which case the residual four-fermion interactions among composites play 
the role of the ETC-induced effective four-fermion interactions between technifermions and SM fermions. 
~\cite{Yamawaki:1982tg}
}
This  old-type ETC model which is based on the scale-up of the QCD 
has some problems; 
Flavor Changing Neutral Current (FCNC) problem~\cite{Dimopoulos:1979es}, 
light pseudo Nambu-Goldstone (NG) bosons and deviation from the LEP precision experiments, 
the so-called $S$ parameter problem~\cite{Peskin:1990zt}.

The FCNC problem and light pseudo NG bosons have already been solved 
by the walking (=scale invariant/conformal) TC~\cite{Holdom:1984sk,Yamawaki:1985zg}
with the TC gauge coupling almost constant or running very slowly, 
which was shown to  develop a large anomalous dimension  $\gamma_m \simeq 1$~\cite{Yamawaki:1985zg}. 
(For reviews see Ref.~\cite{Miransky:vk,Yamawaki:1996vr, Hill:2002ap}.)
It was suggested~\cite{Appelquist:1996dq,Miransky:1996pd} 
that a realistic walking/conformal gauge theory 
is given by the large $N_f$ QCD, a jargon for a version of the ``QCD'' 
with $N_c$ colors and many massless flavors $N_f (\gg N_c)$, 
in which the two-loop beta function possesses the Banks-Zaks infrared fixed point (BZ-IRFP) $\alpha_*$ 
for large $N_f$~\cite{Banks:1981nn}.~\footnote{
There is another possibility to have the Banks-Zaks IR fixed point without 
large $N_f$ by introducing higher dimensional representation~\cite{Hong:2004td}.
} 
such that $N_f^* < N_f < 11 N_c/2$, with $N_f^* \simeq 8.01$ for $N_c=3$. 
Looking at the region $0< \alpha < \alpha_*$, we note that  $ \alpha_* \searrow 0$ when $N_f  \nearrow 11 N_c/2$, 
and hence there exists a certain region $(N_f^* <) \, N_f^{\rm cr}< N_f < 11 N_c/2 $ (``conformal window'') 
such that   $ \alpha_*  < \alpha^{\rm crit}$,  
where $\alpha^{\rm crit} $ is the critical coupling for the chiral symmetry breaking 
and hence the chiral symmetry gets restored $\VEV{\psi}=0$ in this region. 
Here $\alpha^{\rm crit} $ may be evaluated as 
$\alpha^{\rm crit} =\pi/3 C_2(F)$ in the ladder approximation, 
in which case  we have $N_f^{\rm cr} \simeq 4 N_c$~\cite{Appelquist:1996dq} 
\footnote{
In the case of  $N_c=3$, this value $N_f^{\rm cr} \simeq 4 N_c =12$ is somewhat different 
from the lattice value~\cite{Iwasaki:2003de}, $6<N_f^{\rm cr}<7$.
}.
When applied to TC, we set $\alpha_*$ slightly larger than $\alpha^{\rm crit}$ 
(slightly outside of the conformal window),  with the running coupling becoming
slightly larger than the critical coupling in the infrared region, 
we have a condensate or the dynamical mass of the technifermion $m_{\rm TC}$ which is much smaller than
the intrinsic scale of the theory $\Lambda_{\rm TC} (\gg m_{\rm TC} )$. 
In a wide region $m _{\rm TC} < \mu < \Lambda_{\rm TC} $ the coupling is walking due to the BZ-IRFP 
and the theory develops a large anomalous dimension $\gamma_m \simeq 1$ 
and enhanced condensate $\VEV{\psi}|_{\Lambda_{\rm TC}} 
\sim \Lambda_{\rm TC}  m_{\rm TC} ^2$ at the scale of $\Lambda_{\rm TC}$ 
which is identified with the ETC scale $\Lambda_{\rm TC} =\Lambda_{\rm ETC}$.

As to the $S$ parameter, 
it is rather difficult at this moment to draw a definite conclusion in the walking/conformal TC, 
since 
there is no reliable non-perturbative calculation of the $S$ parameter in the walking/conformal TC 
which is strongly coupled and has no simple scale-up of the known dynamics like QCD.
Actually, it was argued ~\cite{Appelquist:1998xf} that 
the $S$ parameter is suppressed in walking/conformal theories with $\gamma_m \simeq 1$. 
Straightforward computation of the $S$ parameter was also performed, 
based on the ladder Schwinger-Dyson (SD) equation and Bethe-Salpeter equation, 
which indicates decreasing tendency in the walking/conformal regime~\cite{Harada:2005ru}.
Recently, 
a reduction of the $S$ parameter was further claimed~\cite{Hong:2006si} 
in a version of the holographic QCD 
with deformation of the replacement of the anomalous dimension $\gamma_m \simeq 0$ 
by $\gamma_m \simeq 1$ based on  AdS/CFT correspondence.

Therefore, 
it would be worth engineering an ETC model building based on the large $N_f$ QCD
near the conformal window as a walking/conformal TC.
There have been such an attempt~\cite{Appelquist:1993sg,Appelquist:2003hn} 
based on a generalized version of Most Attractive Channel (MAC) analysis, 
which claimed  reasonable phenomenological result.
However, it  does not seem to explain a large mass of the top quark (top-bottom splitting)
in a way consistent with the $\rho/T$ parameter constraint from the LEP precision experiments, 
the problem being more serious for the walking/conformal TC.~\cite{Chivukula:1988qr}

Such a large top quark mass may be explained by the top quark condensate, 
or top-mode standard model~\cite{Miransky:1988xi,Nambu89,Marciano89,BHL90}.
So, it would be natural to seek a model 
which accommodates top quark condensate into an explicit scheme of ETC.
After ETC breaking,  it would yield a low energy effective theory something resembling 
the topcolor-assisted technicolor (TC2)~\cite{Hill:1994hp,Hill:1991at} as a variant of
the flavor--universal TC2\,~\cite{Lane:1995gw} 
rather than of others: 
``classic TC2''\,~\cite{Hill:1994hp} and ``type I\kern-.1em I TC2''\,~\cite{Hill:1991at}.  

In this paper we experiment such an explicit ETC model incorporating straightforwardly 
the flavor-universal coloron type topcolor.
Apart from the topcolor sector, 
the model is the same as that of Ref. \cite{Appelquist:1993sg,Appelquist:2003hn}; 
Namely, the ETC gauge group $SU(5)_{\rm ETC}$ 
contains one-family $SU(2)_{\rm TC}$ walking/conformal technicolor 
and $SU(3)$ horizontal gauge symmetry for three families of the SM fermions. 
Since the flavor-universal coloron acts  like  $SU(3)_{\rm QCD}$ gluon on the quarks and techniquarks, 
the model is the same as that of  Ref. \cite{Appelquist:1993sg,Appelquist:2003hn} concerning the ETC sector. 
It was shown in Ref. \cite{Appelquist:1993sg,Appelquist:2003hn} that 
$SU(5)_{\rm ETC}$  can break down successively to $SU(4)_{\rm ETC}$, $SU(3)_{\rm ETC}$ 
and eventually to  $SU(2)_{\rm TC}$, the walking/conformal TC,
by the Most Attractive Channel (MAC) analyses of the dynamics of ETC 
and $SU(2)_{\rm HC}$ hypercolor. However,  actual criticality conditions were not fully
considered for the ETC breakings and 
all three ETC breaking scales $\Lambda_1>\Lambda_2> \Lambda_3$ were thus treated as free parameters. 
Here we impose criticality condition for every step of
the ETC breakings, based on the ladder SD equation,  and find that 
the ETC breaking scales are no longer
free parameters but actually are determined once we fix the scale $\Lambda_1$
for the initial breaking $SU(5)_{\rm ETC} \rightarrow SU(4)_{\rm ETC} $  as an input. 
It turns out that the scales for the
second and third stages of the ETC breakings, $\Lambda_2,\Lambda_3$, 
are much higher than in Ref. \cite{Appelquist:1993sg,Appelquist:2003hn} 
and hence yield extremely small mass ${\cal O} (10^{-1}\,\GeV)$ 
for the third generation fermions characterized by $\Lambda_3$, 
even though the TC condensate is enhanced by the large anomalous dimension of the walking/conformal TC. 

Then we consider a straightforward extension of such an ETC  model (``top-mode ETC'') 
so as to incorporate the flavor-universal $SU(3)_1\times SU(3)_2$ topcolor, 
by simply replacing the  $SU(3)_{\rm QCD}$ of 
both techniquarks and quarks in the conventional ETC model~\cite{Appelquist:1993sg,Appelquist:2003hn} 
by a single factor group $SU(3)_1$ of the topcolor. 
The topcolor  breaks down to $SU(3)_{\rm QCD}$ color symmetry, 
giving rise to the coloron mass $M_{\rm C}$ and 
 yields effective four-fermion interaction among quarks which is tuned close to the critical coupling. 
This is a new ingredient absent in the conventional ETC model~\cite{Appelquist:1993sg,Appelquist:2003hn}.
Then the broken ETC-induced four-fermion interactions prefer 
the third generation quark condensate and the $U(1)_Y$ interaction
further prefers the top quark condensate to the bottom condensate.
Accordingly,  the top quark condensate takes place 
and the top quark acquires the mass ${\hat m}_t$ from the top quark condensate which is expected to dominate 
the ETC-induced mass $m^{(0)}_t \, (\ll  {\hat m}_t) $: $m_t ={\hat m}_t +m^{(0)}_t$. 
Thus the model appears to solve the third generation mass problem 
of the ETC model of Ref.~\cite{Appelquist:1993sg,Appelquist:2003hn}. 

However, since we require the topcolor $SU(3)_1$ is near criticality and thus is very strong, 
the techniquarks, 
carrying the same strong topcolor as well as the equally strong walking/conformal technicolor, 
are forced to condense at the scale much larger than the weak scale, a disaster. 

Then we consider an alternative (``twisted model''), 
namely a version of TC2 with the flavor-universal coloron type topcolor 
$SU(3)_1$ for the quarks but $SU(3)_2$ for the techniquarks. 
Although explicit ETC model building of this type of TC2 is rather complicated 
and not available at this moment, 
we hope that some larger picture would make it. 
Once we admit a possibility that 
such an ``ETC'' gives rise to hierarchical scales to discriminate the generations of quarks/leptons 
as in the ordinary ETC, 
we can reproduce the quark/lepton masses including the third generation: 
Only the top quark has mass from the top quark condensate 
as well as from the TC condensate via ``ETC''-induced four-fermions,
while other fermions acquire mass only from the latter.

A novel feature we find is the large enhancement of the ``ETC''-induced mass which acts like the bare mass
explicitly breaking the chiral symmetry of the topcolor sector. 
Here we recall the fact~\cite{Kikukawa:1989fw,Kondo:1992sq} that 
the  Nambu-Jona-Lasinio (NJL) -type four-fermion interactions 
develop large anomalous dimension $\gamma_m \simeq 2$ in 
{\it both broken and symmetric phases} 
as far as the couplings are close to the criticality
\footnote{
Although the pure NJL in four dimensions is not renormalizable 
and the concept of the anomalous dimension is not well-defined, 
the present case of the broken topcolor with QCD coupling is actually the gauged NJL model 
which is renormalizable in both broken and symmetric phases 
when $A=24/(33-2N_f) >1$~\cite{ Kondo:1992sq,Kondo:1991yk} and 
in fact in our case $A =8/7>1$ for $N_f=6$ ($\mu<m_{\rm TC}$) 
and $A =24/13>1$ for $N_f=10$ ($m_{\rm TC}<\mu<M_{\rm C}$).
In this case the anomalous dimension has a log correction to 2: 
$Z_m^{-1}\simeq (\Lambda/\mu)^2 [\ln(\Lambda/\Lambda_{\rm QCD})/\ln(\mu/\Lambda_{\rm QCD})]^{-A/2}$. 
See Ref.\cite{Yamawaki:1996vr}.
\label{foot-RP}
}.  
This implies great amplification of the bare mass
$m^{(0)}$ defined at the cutoff scale $\Lambda$ 
to the renormalized mass $m (\mu)$ defined at the lower scale $\mu (\ll \Lambda)$ 
(usually the dynamical mass of the fermion): 
$m (\mu) =Z^{-1}_m\, m^{(0)}$ with $Z^{-1}_m = (\Lambda/\mu)^{\gamma_m} \simeq (\Lambda/\mu)^2$. 
In the case at hand, 
while the unbroken topcolor develops rather small anomalous dimension for the region $M_{\rm C}<\mu<\Lambda_3$, 
the broken topcolor for the region ${\hat m}_t (\simeq m_t) < \mu < M_{\rm C}$ 
yields a large anomalous dimension to the top and bottom quarks 
whose four-fermion couplings 
(in the broken phase for top, while in the symmetric  phase for bottom) 
are both close to the criticality. 
This amplifies the bare mass 
by the factor $Z_m^{-1} \simeq (M_{\rm C}/m_t)^2[\ln(M_{\rm C}/\Lambda_{\rm QCD})/\ln(m_t/\Lambda_{\rm QCD})]^{-A/2}$ 
(with log correction due to QCD) 
which would be $Z_m^{-1}  > 500$ 
for e.g., $M_{\rm C} >4\, {\rm TeV}$. 
This easily realizes renormalized ETC-induced mass $\simeq 5 \,\GeV$ for the top and bottom, 
This gives us the renormalized ETC-induced mass for the top and bottom 
even when the typical ETC-induced bare mass at the scale of ETC breaking is very tiny: 
$\sim {\cal O} (100 \,\MeV)$. 
Then the main portion of the top mass coming from the top quark condensate is 
${\hat m}_t = m_t - 5\,\GeV=167\,\GeV$ so that we can reproduce the physical top mass: 
$m_t \simeq 167 + 5=172 \,\GeV$. 
Other quarks having four-fermion couplings somewhat smaller than the criticality 
will have the anomalous dimension smaller than that of the third generation, 
so that the enhancement would be much smaller.
  
So we hope that 
it can be the first step toward 
constructing a realistic model incorporating both technicolor and topcolor.
 
As a novel effect of the large anomalous dimension $\gamma_m \simeq 2$ of the topcolor dynamics mentioned above, 
we find the upper bound of the top-pion mass $m_{\pi_t} < 70\,\GeV$ 
which is conservative estimate and is universal to the generic TC2 not restricted to our model setting. 
The top-pion appears in the generic TC2 model\cite{Hill:1994hp,Hill:1991at}, 
since both the top quark condensate and technifermion condensate break respective global symmetries, 
which results in two kinds of three  Nambu-Goldstone (NG) bosons. 
Three of them (mainly boundstates of technifermions) are absorbed into $W,Z$  bosons as usual 
and the rest three are pseudo NG bosons (mainly boundstates of top and bottom). 
The top-pion  mass may be estimated by the Dashen formula~\cite{Dashen:1969eg} 
$m^2_{\pi_t} f^2_{\pi_t}= m_t^{(0)} \VEV{t}$, 
where $ m^{(0)}_t$ is the ETC-induced top mass acting as the bare mass. 
The decay constant $f_{\pi_t}$ may be calculated by the Pagels-Stokar formula~\cite{Pagels:1979hd} 
for $f^2_{\pi_t}$ which is evaluated in exactly the same way 
as in the original top quark condensate paper~\cite{Miransky:1988xi} 
where the mass function is not a constant but logarithmically damping due to the QCD correction. 
It reads: $f^2_{\pi_t} = [N_c ({\hat m}_t)^2/(8 \pi^2)] \cdot F(M_{\rm C},{\hat m}_t)$, 
where the function $F$ is $\ln (M_{\rm C}^2/({\hat m}_t)^2)$ in the pure NJL model 
but is modified to a certain function in the gauged NJL model 
which is finite in the limit $M_{\rm C} \rightarrow \infty$, 
reflecting the renormalizability of the gauged NJL model. 

Now, the crucial point of our estimate is that the right-hand side of the Dashen formula is 
{\it renormalization-point independent}: 
$m^{(0)}_t(M_{\rm C}) \cdot \VEV{t}|_{M_{\rm C}} = m^{(0)}_t({\hat m}_t) \cdot \VEV{t}|_{{\hat m}_t}$.
Using the estimation $\VEV{t}|_{{\hat m}_t}= N_c ({\hat m}_t)^3/(4\pi^2)$, 
we have 
$m^2_{\pi_t} = m^{(0)}_t \VEV{t}/ f^2_{\pi_t} 
= 2 m^{(0)}_t({\hat m}_t) \cdot {\hat m}_t/ F(M_{\rm C},{\hat m}_t) 
= 2 x (m_t -x) /  F(M_{\rm C},{\hat m}_t = m_t - x)$, 
where we have written $x \equiv m^{(0)}_t({\hat m}_t)$.
The gross structure of this expression is essentially determined by the factor $2 x (m_t -x)$ 
which has a maximum value $m_t^2/2$ at $x=m_t/2$, and the function $F(M_{\rm C},{\hat m}_t=m_t-x)$, 
numerically similar to $\ln (M_{\rm C}^2/ (m_t-x)^2)$, 
is a slowly increasing function as $M_{\rm C}$ (not diverging, though) 
and hence the lowest  possible $M_{\rm C}$ gives the upper bound of $m_{\pi_t} ^2$. 
A model-independent lower limit of the mass of the flavor-universal coloron 
(as in the class of models considered in this paper) 
$M_{\rm C}$ is $M_{\rm C} > 837\,\GeV$~\cite{Bertram:1998wf}, 
which yields an upper bound $m^2_{\pi_t} < (60\,\GeV)^2$ at $x \simeq m_t/2$. 
On the other hand, 
flavor-non-universal coloron mass bound is somewhat weaker $M_{\rm C} > 450 \,\GeV$ 
which implies the upper bound $m^2_{\pi_t} < (70\,\GeV)^2$.
The latter is a very conservative upper bound universal to generic model of TC2 not restricted to specific TC2 model,
since in the generic TC2 model the actual mass bound of the coloron is 
$M_{\rm C}/\cot \theta > 837\,\GeV$ and $M_{\rm C}/\cot \theta > 450\,\GeV$ 
for flavor-universal and flavor-non-universal cases, respectively, 
with the condition that  $\cot \theta >4$ in order to trigger the top quark condensate~\cite{Bertram:1998wf}.
 
The paper is organized as follows:
In Sec.~\ref{A-TMETC},  we present an ETC model incorporating the topcolor (Top-Mode ETC). 
The successive ETC breakings through MAC analyses 
is shown in the same way as the model without topcolor.
In Sec.~\ref{crit-sETC}, 
we discuss criticality conditions for the ETC breakings 
and estimate the breaking scales, 
based on the ladder SD equation analysis. We find that without topcolor 
the model yields only a small mass on the order of 
${\cal O} ( 0.1 \,\GeV$)  to the third generation.
In Sec.~\ref{crit-EWSB}, we study the effect of the topcolor in the form of 
the gauged NJL model, the effective four-fermion interactions due to
the broken ETC and topcolor interactions with the SM gauge interactions. 
Based on the phase structure of the gauged NJL model, 
we discuss the criticality conditions under which top quark is the only quark to condense. 
Then it is shown that 
the techniquark condensate, 
which is triggered by the combined effects of the technicolor and the topcolor, 
is  very large as far as we require  
the topcolor coupling is large enough to trigger the top quark condensate. 
In Sec.~\ref{fullU-TC2} we argue an alternative model of TC2, 
where we roughly estimate third generation quark masses  
and emphasize that the large anomalous dimension $\gamma_m \simeq 2$ 
of the broken topcolor as well as the large anomalous dimension of the walking/conformal TC 
is crucial to give the large enhancement for the third generation quark masses.
In Sec.~\ref{top-pion mass} we give a new estimate of the mass of top-pion 
as a novel effect of the large anomalous dimension of the broken topcolor.  
We find a rather small upper bound of the top-pion, 
which is universal to the generic TC model. 
Sec.~\ref{sum} is devoted to summary and discussions. 

\section{A Top Mode Extended Technicolor Model}
\label{A-TMETC}
\subsection{The Model}
\label{pro-TMETC}

We use a typical one-family TC model~\cite{Hill:2002ap}
with $N_f = 8$ technifermions and with $SU(2)_{\rm TC}$ as a TC gauge group, 
which is a walking/conformal TC near the edge of 
the conformal window $N_f \sim 4 N_c$~\cite{Appelquist:1996dq} evaluated 
in the ladder approximation. 
The simplest ETC model would be $SU(5)_{\rm ETC}$ 
which accommodates $SU(2)_{\rm TC}$ one-family technifermions and three families of quarks and leptons. 
Following Ref.~\cite{Appelquist:2003hn}, we introduce $SU(2)_{\rm HC}$ in order that 
the $SU(5)_{\rm ETC}$ successively breaks down 
as $SU(5)_{\rm ETC} \to SU(4)_{\rm ETC} \to SU(3)_{\rm ETC} \to SU(2)_{\rm TC}$.
Although this model is the same as that of Ref.~\cite{Appelquist:2003hn} as to the ETC sector, 
we introduce the universal coloron-type topcolor symmetry $SU(3)_1 \times SU(3)_2$ 
which breaks down to $SU(3)_{\rm QCD}$.
Note that the universal coloron does not affect the ETC sector.   

Particle contents in this model is listed in Table.~\ref{particle-content}. 
${\cal Q},{\cal U},{\cal D},{\cal L},{\cal E}$ include one-family technifermions 
and three-generation quarks and leptons :
\beq
{\cal Q}_L=( Q^a\,,\,q_3\,,\,q_2\,,\,q_1 )^T_L\,,\,
{\cal L}_L=( L^a\,,\,l_3\,,\,l_2\,,\,l_1 )^T_L\,,\,
{\cal U}_R=( U^a\,,\,t\,,\,c\,,\,u)^T_R\,,\cdots\,,
\eeq
where $a=1,2$ is TC indices, $q_i(l_i)$ represents $i$-th generation 
$SU(2)_L$-doublet quark(lepton), 
$Q^a= ( U^a \,,\,D^a)^T$ and $L^a= ( N^a \,,\,E^a)^T$ are techniquarks and technileptons, respectively. 
Additional fermions $\psi_R\,,\,\psi'_R$ 
participate in the desirable ETC breaking as will be discussed in detail in Sec.~\ref{SB-ETC}, 
while $\omega_R$ contributes only to the running behavior of the $SU(2)_{\rm HC}$ gauge coupling, 
and the largest possible number of $\omega$ is $N_\omega \leq 10$ 
in order to keep the asymptotic freedom 
($N_\omega = 2$ in Ref.~\cite{Appelquist:2003hn}, while we shall take $N_\omega = 10$).
The condensation of ``effective Higgs'' field $\Phi$ breaks the topcolor symmetry to $SU(3)_{\rm QCD}$ symmetry.
\begin{table}
\begin{center}
\begin{tabular}{| c | c | c | c | c | c || c |
}
\hline
field & $ SU(5)_{\rm ETC} $ & $ SU(3)_1$  & $SU(3)_2$ & $SU(2)_L$ & $U(1)_Y$ 
& $SU(2)_{\rm HC}$ 
\\
\hline 
${\cal Q}_L$ & 5 & 3 & 1 & 2 & $1/6$ & 1 
\\
${\cal U}_R$ & 5 & 3 & 1 & 1 & $2/3$ & 1 
\\
${\cal D}_R$ & 5 & 3 & 1 & 1 & $-1/3$ & 1 
\\ 
\hline
${\cal L}_L$ & 5 & 1 & 1 & 2 & $-1/2$ & 1 
\\
${\cal E}_R$ & 5 & 1 & 1 & 1 & $-1$ & 1 
\\ 
\hline
\, & \, & \, & \, & \, & \, & \, 
\\[-2.5ex] 
\hline
$\psi_R$ & 10 & 1 & 1 & 1 & 0 & 2  
\\
$\psi'_R$ & $\overline{10}$ & 1 & 1 & 1 & 0 & 1 
\\
$\omega_R$ & 1 & 1 & 1 & 1 & 0 & 2 
\\
\hline 
\, & \, & \, & \, & \, & \, & \, 
\\[-2.5ex] 
\hline 
\rule[0pt]{0pt}{12pt}
$\Phi$ & 1 & 3 & $\overline{3}$ & 1 & 0 & 1 \\
\hline
\end{tabular}
\caption{Particle contents.
         This table without $\Phi$ 
         is the same as Ref.~\cite{Appelquist:2003hn} 
         beside that $SU(3)_{\rm QCD}$ is replaced by $SU(3)_1 \times SU(3)_2$.
         $\Phi$ is the scalar field and 
         other fields are the fermion fields.
         \label{particle-content}}
\end{center}
\end{table}%

Apart from the ETC group, 
this is the same as the  flavor-universal TC2~\cite{Lane:1995gw} 
in the sense that all quarks (techniquarks as well) have the same charge 
under the topcolor symmetry $ SU(3)_1 \times SU(3)_2$, 
but for simplicity we do not introduce the additional (strong) $U(1)'$ 
which in the flavor-universal TC2 models 
is coupled only to the third generation quark to trigger the top condensate.
Instead, the ETC gauge interaction in our case discriminates the third generation quarks 
from others near the criticality of the strong coloron interaction. 
Once the third generation is selected, 
top is distinguished from bottom by the usual $U(1)_Y$ interaction in the Standard Model near the criticality.
\footnote{
This requires some fine tuning 
which may be avoided by introducing extra strong $U(1)'$ as in the flavor--universal TC2 models. 
Although the conventional $U(1)'$ to distinguish the third generation from others 
may not straightforwardly be incorporated into the ETC model, the flavor-universal $U(1)'$~\cite{Braam:2007pm} is
an interesting possibility. We shall study such a case in Sec.\ref{fullU-TC2}.  
}

Before discussing detailed dynamics, we here note that the desired ETC breaking: 
$SU(5)_{\rm ETC} \to SU(4)_{\rm ETC} \to SU(3)_{\rm ETC} \to SU(2)_{\rm TC}$
and topcolor breaking: $SU(3)_1 \times SU(3)_2 \to SU(3)_{\rm QCD}$ 
can be realized through the (electroweak singlet) ``effective Higgs" fields  
\beq
H^{\rm ETC}_1 &\sim& (5,1,1) 
\,\,(\text{under}\,\,SU(5)_{\rm ETC} \times SU(3)_1 \times SU(3)_2) \,,\nonumber\\
H^{\rm ETC}_2 &\sim& (\overline{4},1,1) 
\,\,(\text{under}\,\,SU(4)_{\rm ETC} \times SU(3)_1 \times SU(3)_2) \,,\nonumber\\
H^{\rm ETC}_3 &\sim& (\overline{3},1,1) 
\,\,(\text{under}\,\,SU(3)_{\rm ETC} \times SU(3)_1 \times SU(3)_2) \,,\nonumber\\
\Phi &\sim& (1,3,\overline{3}) 
\,\,(\text{under}\,\,\,SU(2)_{\rm TC}\,\,\times SU(3)_1 \times SU(3)_2)\,.
\eeq
Here ``effective Higgs" fields $H^{\rm ETC}_1\,,\,H^{\rm ETC}_2\,,\, H^{\rm ETC}_3$ break 
$SU(5)_{\rm ETC}$, $SU(4)_{\rm ETC}$ and  $SU(3)_{\rm ETC}$ successively through hierarchical
condensates $\Lambda_1 \gg \Lambda_2 \gg \Lambda_3$ 
where $\Lambda_i = \vev{H^{\rm ETC}_i}$ .
$\Phi$ breaks $SU(3)_1 \times SU(3)_2$ through $\vev{\Phi}$ 
which we take $\Lambda_3 > \Lambda_{\rm C} = \vev{\Phi}$. 

Although a scenario of dynamically producing such ``effective Higgs'' fields 
for the successive ETC breakings was given in Ref~\cite{Appelquist:1993sg,Appelquist:2003hn}, 
it was left unclear whether or not the criticality conditions for 
each step of ETC breakings can really be met.
Here, we shall explicitly examine the criticality conditions 
and find that in order for the $SU(5)_{\rm ETC} \to SU(4)_{\rm ETC}$ breaking takes place, 
the $SU(5)_{\rm ETC}$ coupling at the scale of $\Lambda_1 (= 1000\,\TeV)$ 
should be much larger than that of Ref~\cite{Appelquist:1993sg}. 
Once $SU(5)_{\rm ETC}$ breaking takes place due to the coupling this large 
the desired successive breakings 
$SU(5)_{\rm ETC} \to SU(4)_{\rm ETC} \to SU(3)_{\rm ETC} \to SU(2)_{\rm TC}$ 
are actually realized.
However, we shall demonstrate that 
the lowest ETC breaking scale $\Lambda_3$ is determined to be very large ; 
$\Lambda_3 \simeq 700\,\TeV\,(N_\omega=2)$ and $\Lambda_3 \simeq 360\,\TeV\,(N_\omega=10)$. 
Even including ambiguity of critical value  
of the ladder SD equation up to $30\%$ for the $SU(5)_{\rm ETC}$, 
it can only be $\Lambda_3 \gtrsim 150\,\TeV$, 
so that the third generation mass should be at most in order of ${\cal O}(10^{-1}\,\GeV)$  
even in the walking/conformal TC with $\gamma_m \simeq 1$ within the framework of only ETC without topcolor.

Now to a model with topcolor. 
Due to $\vev{\Phi}=\Lambda_{\rm C}$ the topcolor symmetry $SU(3)_1 \times SU(3)_2$ 
is spontaneously broken down to $SU(3)_{\rm QCD}$. 
We are left with strongly coupled effective four fermion interaction 
which, combined with broken ETC and $U(1)_Y$ gauge interactions near the criticality, 
triggers the top quark condensate giving rise to the main part of the top quark mass 
$m^{\rm topC}_t \simeq 170\,\GeV$. 
In the case of $N_\omega=10$, 
we have $\Lambda_2 \simeq 850\,\TeV\,,\,\Lambda_3 \simeq 360\,\TeV$ for $\Lambda_1 = 1000\,\TeV$.
If the TC dynamics is near critical dynamics 
then the ETC driven mass of the third generation $m^{\rm ETC}_{t,b} \simeq 10^{-1}\,\GeV$ 
which is regarded as the bare mass at $\Lambda_3$ in the topcolor dynamics 
and is expected to be amplified to $m^{\rm ETC}_{t,b} \simeq 5\,\GeV$ at a scale of top mass 
by the anomalous dimension of NJL-type $\gamma_m \simeq 2$ 
for the quark bilinear operator due to broken topcolor dynamics. 
Then we would have $m_t = m^{\rm topC}_t + m^{\rm ETC}_t \simeq 175\,\GeV$ and
$m_b = m^{\rm ETC}_b \simeq 5\,\GeV$.
However there is a serious drawback in this model: Combined effects of the technicolor and topcolor are
very strong, which trigger the techniquark condensate at much larger scale than the weak scale. Then the
$f_{\pi_T}$ does not satisfy the basic requirement of the model setting 
$4f^2_{\pi_T} \simeq 4 (110\,\GeV)^2 = (246\,\GeV)^2 - f^2_{\pi_t} \,,\,f^2_{\pi_t} \simeq (100\,\GeV)^2$.
We shall discuss a possible way out in the latter section. 

\subsection{MAC analysis of Successive ETC breakings}
\label{SB-ETC}

In this section we review the MAC analysis of $SU(5)_{\rm ETC}$ with $SU(2)_{\rm HC}$ 
following Ref.~\cite{Appelquist:1993sg,Appelquist:2003hn},
postponing our own discussions on the criticality of the MAC binding strength 
to Sec.\ref{crit-sETC}\footnote{
Since the universal coloron type topcolor acts 
in the same way as the $SU(3)_{\rm QCD}$ in the discussions of ETC breakings, 
discussions in Sec.\ref{SB-ETC} and Sec.\ref{crit-sETC} 
apply to both the model of Ref~\cite{Appelquist:1993sg,Appelquist:2003hn} and ours.
}.

The successive ETC breaking may be realized by 
$\psi_R\,,\,\psi'_R$ (topcolor and EW singlets) in Table.~\ref{particle-content} 
and
the $SU(2)_{\rm HC}$ gauge interaction.
Let us define 
$\Delta C_2(r_1 \times r_2 \to r_3) \equiv C_2(r_1) + C_2(r_2) - C_2(r_3)$ 
where $C_2(r)$ is a quadratic Casimir operator 
with representation $r$ under each gauge group.
Also, 
$g_{N ({\rm ETC})}$ is $SU(N)_{\rm ETC}$ gauge coupling, 
$\alpha_{N ({\rm ETC})}=g_{N ({\rm ETC})}^2/4\pi$, 
and 
$g_{2 ({\rm HC})}$ is $SU(2)_{\rm HC}$ gauge coupling, 
$\alpha_{2({\rm HC})}=g_{2 ({\rm HC})}^2/4\pi$.
\subsubsection{Realization of $SU(5)_{\rm ETC}$ breaking down to $SU(4)_{\rm ETC}$}
\label{54-AMAC}

First, 
$\psi_R\,,\,\psi'_R$ in Table.~\ref{particle-content} take part in $SU(5)_{\rm ETC}$ breaking system.
The preserving $SU(2)_{\rm HC}$ candidates of condensation are :
\beq
&&(\overline{10},1,1,1,1)_0\times (\overline{10},1,1,1,1)_0 \to (5,1,1,1,1)_0\,,\nonumber\\
&&\hspace*{35ex}: k^{(5,1)}_5=\frac{24}{5}\alpha_{5(\rm ETC)}(\Lambda_1)\,,
\label{5ETC-1}\\
&&(10,1,1,1,2)_0\times (10,1,1,1,2)_0 \to (\overline{5},1,1,1,1)_0\,,\nonumber\\
&&\hspace*{35ex}: k^{(\overline{5},1)}_5=\frac{24}{5}\alpha_{5(\rm ETC)}(\Lambda_1) 
+ \frac{3}{2} \alpha_{2(\rm HC)}(\Lambda_1)\,,
\label{5ETC-2}\\
&&(10,1,1,1,2)_0\times (10,1,1,1,2)_0 \to (\overline{5},1,1,1,3)_0\,,\nonumber\\
&&\hspace*{35ex}: k^{(\overline{5},3)}_5=\frac{24}{5}\alpha_{5(\rm ETC)}(\Lambda_1) 
- 2\alpha_{2(\rm HC)}(\Lambda_1)\,,
\label{5ETC-3}
\eeq 
where representations in the parentheses correspond to  
$(SU(5)_{\rm ETC}$, $SU(3)_1$, $SU(3)_2$, $SU(2)_L$ , $SU(2)_{\rm HC}$ $)_{U(1)_Y}$, 
and each $\Delta C_2$ is
$\Delta C_2(\overline{10}\times \overline{10} \to 5)=
\Delta C_2(10\times 10 \to \overline{5})=24/5$ 
for $SU(5)_{\rm ETC}$ and
$\Delta C_2(2 \times 2 \to 1)=3/2$, 
$\Delta C_2(2 \times 2 \to 3)=-2$ for $SU(2)_{\rm HC}$.
Here $\kappa^{(A,B)}_N$ represents 
a binding strength for each channel 
labeled by the representations ($A$ , $B$) 
of condensation under $(SU(N)_{\rm ETC},SU(2)_{\rm HC})$.
We can see easily 
$k^{(\overline{5},1)}_5>k^{(5,1)}_5>k^{(\overline{5},3)}_5$, 
so that MAC appears to be Eq.(\ref{5ETC-2}) rather than Eq.(\ref{5ETC-1}) and (\ref{5ETC-3}).
However, 
the channel Eq.(\ref{5ETC-2}) is forbidden by the Fermi statistics, 
and hence Eq.(\ref{5ETC-1}) is the MAC 
for the breaking as $SU(5)_{\rm ETC} \to SU(4)_{\rm ETC}$.
The condensation of this channel corresponds to 
$\vev{H^{\rm ETC}_1} = \Lambda_1 \neq 0$ in Sec.~\ref{pro-TMETC}.
Once 
$k^{(5,1)}_5$ exceeds 
the critical binding strength $k_{\crit}$ (as will be discussed in Sec.~\ref{crit-sETC}), 
$SU(5)_{\rm ETC}$ breaks down to $SU(4)_{\rm ETC}$, 
and as a result 
quark/lepton sector is divided below $\Lambda_1$ as 
\beq
{\cal Q}_L \to \left\{ \!\!\!
 \begin{array}{l}
 (1,3,1,2,1)_{1/6} \,\,: q_{1L}=(u,d)_L\\[1.5ex]
 (4,3,1,2,1)_{1/6} \,\,: {\cal Q}'_L
 \end{array}\right.\,
\hspace*{-5ex}
&&
,\,\,
{\cal L}_L \to \left\{ \!\!\!
 \begin{array}{l}
 (1,1,1,1,1)_{-1/2} \,\,: l_{1L}=(\nu_e,e)_L\\[1.5ex]
 (4,1,1,1,1)_{-1/2} \,\,: {\cal L}'_L
 \end{array}\right.
\\[3ex]
{\cal U}_R \to \left\{ \!\!\!
 \begin{array}{l}
 (1,3,1,1,1)_{2/3} \,\,: u_R\\[1.5ex]
 (4,3,1,1,1)_{2/3} \,\,: {\cal U}'_R
 \end{array}\right.\,\!
\hspace*{5ex}
&&
,
\\[3ex]
{\cal D}_R \to \left\{ \!\!\!
 \begin{array}{l}
 (1,3,1,1,1)_{-1/3} \,\,: d_R\\[1.5ex]
 (4,3,1,1,1)_{-1/3} \,\,: {\cal D}'_R
 \end{array}\right.\,\!\!
\hspace*{4ex}
&&
,\,\,
{\cal E}_R \to \left\{ \!\!\!
 \begin{array}{l}
 (1,1,1,1,1)_{-1} \,\,: e_R\\[1.5ex]
 (4,1,1,1,1)_{-1} \,\,: {\cal E}'_R
 \end{array}\right.\, 
\eeq
and the remaining fields of $\psi_R\,,\,\psi'_R$ are  
\beq
\psi_R : \{(4,1,1,1,2)_0\,,\,(6,1,1,1,2)_0\} \,\,,\,\, \psi'_R : (\overline{4},1,1,1,1)_0 \,,
\eeq
where we labeled the representations according to 
 ($SU(4)_{\rm ETC}$, $SU(3)_1$, $SU(3)_2$, $SU(2)_L$, $SU(2)_{\rm HC}$ )$_{U(1)_Y}$.

\subsubsection{Realization of $SU(4)_{\rm ETC}$ breaking down to $SU(3)_{\rm ETC}$}
\label{43-AMAC}

Now, $(4,1,1,1,2)_0\,,\,(6,1,1,1,2)_0\,,\,(\overline{4},1,1,1,1)_0$ 
and the $SU(2)_{\rm HC}$ gauge interaction take part in the $SU(4)_{\rm ETC}$ breaking system.
The candidates of condensation are :
\beq
&&(4,1,1,1,2)_0\times (6,1,1,1,2)_0 \to (\overline{4},1,1,1,1)_0\,,\nonumber\\
&&\hspace*{35ex}
  : k^{(\overline{4},1)}_4=\frac{5}{2}\alpha_{4(\rm ETC)}(\Lambda_2)
  + \frac{3}{2} \alpha_{2(\rm HC)}(\Lambda_2)\,,
\label{4ETC-1}\\
&&(6,1,1,1,2)_0\times (6,1,1,1,2)_0 \to (1,1,1,1,3)_0\,,\nonumber\\
&&\hspace*{35ex}
  : k^{(1,3)}_4=5 \alpha_{4(\rm ETC)}(\Lambda_2) 
  - \frac{1}{2} \alpha_{2(\rm HC)}(\Lambda_2)\,,
\label{4ETC-2}
\eeq 
where the representations were labeled by 
($SU(4)_{\rm ETC}$, $SU(3)_1$, $SU(3)_2$, $SU(2)_L$, $SU(2)_{\rm HC}$)$_{U(1)_Y}$
and 
$\Delta C_2 (4\times 6 \to \overline{4}) = 5/2$, 
$\Delta C_2 (6\times 6 \to 1) = 5$ for $SU(4)_{\rm ETC}$.

The channel in Eq.(\ref{4ETC-1}) is the MAC rather than Eq.(\ref{4ETC-2}) 
if  
\beq
k^{(\overline{4},1)}_4>k^{(1,3)}_4 
\Leftrightarrow
\alpha_{2(\rm HC)}(\Lambda_2) > \frac{5}{4}\alpha_{4(\rm ETC)}(\Lambda_2)
\,,
\label{43MAC}
\eeq
is satisfied at $\Lambda_2$.
The condensation of Eq.(\ref{4ETC-1}) correspond to 
$\vev{H^{\rm ETC}_2} = \Lambda_2 \neq 0$ in Sec.~\ref{pro-TMETC}.
Once 
$k^{(\overline{4},1)}_4$ exceeds  
the critical binding strength $k_{\crit}$ (as will be discussed in Sec.~\ref{crit-sETC}), 
$SU(4)_{\rm ETC}$ breaks down to $SU(3)_{\rm ETC}$, 
and as a result quark/lepton sector is divided below $\Lambda_2$ as 
\beq
{\cal Q}_L \to \left\{ \!\!\!
 \begin{array}{l}
 (1,3,1,2,1)_{1/6} \,\,: q_{1L}=(u,d)_L\\[1.5ex]
 (1,3,1,2,1)_{1/6} \,\,: q_{2L}=(c,s)_L\\[1.5ex]
 (3,3,1,2,1)_{1/6} \,\,: {\cal Q}''_L
 \end{array}\right.\,
\hspace*{-5ex}
&&
,\,\,
{\cal L}_L \to \left\{ \!\!\!
 \begin{array}{l}
 (1,1,1,1,1)_{-1/2} \,\,: l_{1L}=(\nu_e,e)_L\\[1.5ex]
 (1,1,1,1,1)_{-1/2} \,\,: l_{2L}=(\nu_\mu,\mu)_L\,\,\,\,\\[1.5ex]
 (3,1,1,1,1)_{-1/2} \,\,: {\cal L}''_L
 \end{array}\right.
\\[3ex]
{\cal U}_R \to \left\{ \!\!\!
 \begin{array}{l}
 (1,3,1,1,1)_{2/3} \,\,: u_R\\[1.5ex]
 (1,3,1,1,1)_{2/3} \,\,: c_R\\[1.5ex]
 (3,3,1,1,1)_{2/3} \,\,: {\cal U}''_R
 \end{array}\right.\,\!
\hspace*{5ex}
&&
,
\\[3ex]
{\cal D}_R \to \left\{ \!\!\!
 \begin{array}{l}
 (1,3,1,1,1)_{-1/3} \,\,: d_R\\[1.5ex]
 (1,3,1,1,1)_{-1/3} \,\,: s_R\\[1.5ex]
 (3,3,1,1,1)_{-1/3} \,\,: {\cal D}''_R
 \end{array}\right.\,\!\!
\hspace*{4ex}
&&
,\,\,
{\cal E}_R \to \left\{ \!\!\!
 \begin{array}{l}
 (1,1,1,1,1)_{-1} \,\,: e_R\\[1.5ex]
 (1,1,1,1,1)_{-1} \,\,: \mu_R\\[1.5ex]
 (3,1,1,1,1)_{-1} \,\,: {\cal E}''_R
 \end{array}\right.\, 
\eeq
and the remaining fields of $\psi_R\,,\,\psi'_R$ are 
\beq
\psi_R : \{(1,1,1,1,2)_0\,,\,(3,1,1,1,2)_0\} \,\,,\,\, \psi'_R :\{(1,1,1,1,1)_0\,,\,(\overline{3},1,1,1,1)_0\}\,,
\eeq
where we labeled the representations by 
$(SU(3)_{\rm ETC}\,,\,SU(3)_1\,,\,SU(3)_2\,,\,SU(2)_L\,,\,SU(2)_{\rm HC})_{U(1)_Y}$.

\subsubsection{Realization of $SU(3)_{\rm ETC}$ breaking down to $SU(2)_{\rm TC}$}
\label{32-AMAC}

Finally, 
$(1,1,1,1,2)_0\,,\,(3,1,1,1,2)_0\,,\,(1,1,1,1,1)_0\,,\,(\overline{3},1,1,1,1)_0$
and the $SU(2)_{\rm HC}$ gauge interaction 
contributes the $SU(3)_{\rm ETC}$ breaking system.
The candidates of condensation are :
\beq
&&(3,1,1,1,2)_0\times (3,1,1,1,2)_0 \to (\overline{3},1,1,1,1)_0\,,\nonumber\\
&&\hspace*{35ex}
  :k^{(\overline{3},1)}_3
  =\frac{4}{3}\alpha_{3(\rm ETC)}(\Lambda_3)
  + \frac{3}{2} \alpha_{2(\rm HC)}(\Lambda_3)\,,
\label{3ETC-1}\\
&&(3,1,1,1,2)_0\times (\overline{3},1,1,1,1)_0 \to (1,1,1,1,2)_0\,\,,\nonumber\\
&&\hspace*{35ex}
  :k^{(1,2)}_3= 
   \frac{8}{3} \alpha_{3(\rm ETC)}(\Lambda_3)\,,
\label{3ETC-2}
\eeq 
where the representations were labeled by 
($SU(3)_{\rm ETC}$, $SU(3)_1$, $SU(3)_2$, $SU(2)_L$, $SU(2)_{\rm HC}$)$_{U(1)_Y}$
and $\Delta C_2 (3\times 3 \to \overline{3}) = 4/3$, 
$\Delta C_2 (3\times \overline{3} \to 1) = 8/3$ for $SU(3)_{\rm ETC}$.

The channel in Eq.(\ref{3ETC-1}) is the MAC rather than Eq.(\ref{3ETC-2}) 
if  
\beq
k^{(\overline{3},1)}_3>k^{(1,2)}_3 
\Longleftrightarrow
\alpha_{2(\rm HC)}(\Lambda_3) > \frac{8}{9}\alpha_{3(\rm ETC)}(\Lambda_3)\,,
\label{32MAC}
\eeq
is satisfied at $\Lambda_3$.
The condensation of Eq.(\ref{3ETC-1}) correspond to 
$\vev{H^{\rm ETC}_3} \neq 0$ in Sec.~\ref{pro-TMETC}.

Once 
$k^{(\overline{3},1)}_4$ exceeds 
the critical binding strength $k_{\crit}$ (as will be discussed in Sec.~\ref{crit-sETC}), 
$SU(3)_{\rm ETC}$ breaks down to $SU(2)_{\rm TC}$, 
and as a result, quark/lepton sector is divided below $\Lambda_3$ as 
\beq
{\cal Q}_L \to \left\{ \!\!\!
 \begin{array}{l}
 (1,3,1,2,1)_{1/6} \,\,: q_{1L}=(u,d)_L\\[1.5ex]
 (1,3,1,2,1)_{1/6} \,\,: q_{2L}=(c,s)_L\\[1.5ex]
 (1,3,1,2,1)_{1/6} \,\,: q_{3L}=(t,b)_L\\[1.5ex]
 (2,3,1,2,1)_{1/6} \,\,: Q^a_L = (U^a,D^a)_L
 \end{array}\right.\,
\hspace*{-5ex}
&&
,\,\,
{\cal L}_L \to \left\{ \!\!\!
 \begin{array}{l}
 (1,1,1,1,1)_{-1/2} \,\,: l_{1L}=(\nu_e,e)_L\\[1.5ex]
 (1,1,1,1,1)_{-1/2} \,\,: l_{2L}=(\nu_\mu,\mu)_L\\[1.5ex]
 (1,1,1,1,1)_{-1/2} \,\,: l_{2L}=(\nu_\tau,\tau)_L\\[1.5ex]
 (2,1,1,1,1)_{-1/2} \,\,: L^a_L=(N^a,E^a)_L
 \end{array}\right.
\\[3ex]
{\cal U}_R \to \left\{ \!\!\!
 \begin{array}{l}
 (1,3,1,1,1)_{2/3} \,\,: u_R\\[1.5ex]
 (1,3,1,1,1)_{2/3} \,\,: c_R\\[1.5ex]
 (1,3,1,1,1)_{2/3} \,\,: t_R\\[1.5ex]
 (2,3,1,1,1)_{2/3} \,\,: U^a_R
 \end{array}\right.\,\!
\hspace*{8ex}
&&
,
\\[3ex]
{\cal D}_R \to \left\{ \!\!\!
 \begin{array}{l}
 (1,3,1,1,1)_{-1/3} \,\,: d_R\\[1.5ex]
 (1,3,1,1,1)_{-1/3} \,\,: s_R\\[1.5ex]
 (1,3,1,1,1)_{-1/3} \,\,: b_R\\[1.5ex]
 (2,3,1,1,1)_{-1/3} \,\,: D^a_R
 \end{array}\right.\,\!\!
\hspace*{7ex}
&&
,\,\,
{\cal E}_R \to \left\{ \!\!\!
 \begin{array}{l}
 (1,1,1,1,1)_{-1} \,\,: e_R\\[1.5ex]
 (1,1,1,1,1)_{-1} \,\,: \mu_R\\[1.5ex]
 (1,1,1,1,1)_{-1} \,\,: \tau_R\\[1.5ex]
 (2,1,1,1,1)_{-1} \,\,: E^a_R
 \end{array}\right.\, 
\eeq
and the remaining fields of $\psi\,,\,\psi'$ are 
\beq
2&\times&(1,1,1,1,2)_0\,,\\
2&\times&(1,1,1,1,1)_0\,,\\
N^a_R &\equiv &(2,1,1,1,1)_0
\eeq
where the representations were labeled by 
($SU(2)_{\rm TC}$, $SU(3)_1$, $SU(3)_2$, $SU(2)_L$, $SU(2)_{\rm HC}$)$_{U(1)_Y}$. 
We identify $(2,1,1,1,1)_0$-field with right-handed techni-neutrino, 
so that technilepton condensation preserves the custodial $SU(2)$, 
$\langle \overline{N}_R N_L \rangle = \langle \overline{E}_R E_L \rangle$. 

As we noted before, 
$\omega_R$ contributes only to the running behavior of the $SU(2)_{\rm HC}$ gauge coupling, 
and the largest possible number of $\omega$ is $N_\omega \leq 10$ 
in order to keep the asymptotic freedom. 
As $N_\omega$ increases, the hierarchy among $\Lambda_i$s becomes large. 
As we discuss in Sec.\ref{crit-sETC}, 
if we take $N_\omega = 2$ as Ref.~\cite{Appelquist:2003hn}, 
all $\Lambda_i$s become nearly degenerate. 
We shall take the largest possible value $N_\omega = 10$ 
in order to maximize the hierarchy.

Since the $SU(2)_{\rm HC}$ is confined at a scale near $\Lambda_3$, 
after ETC gauge group breaks down to TC gauge group, 
we have ordinary SM quarks/leptons and one family technifermions 
except for the $SU(3)_1 \times SU(3)_2$ instead of $SU(3)_{\rm QCD}$.

\section{Criticality of the successive ETC breakings}
\label{crit-sETC}

Now we come to the discussions on the criticality of the MAC 
identified in the previous section.
By taking account of the criticality condition 
and the running effect of the ETC gauge couplings at each breaking stage, 
we obtain definite value of $\Lambda_2$ and $\Lambda_3$ 
once $\Lambda_1(=1000\,\TeV)$ is fixed as an input, 
the lowest allowed scale from the $K_0 \overline{K}_0$-mixing.
This is in contrast to Ref~\cite{Appelquist:1993sg,Appelquist:2003hn}
which did not impose the criticality condition 
for the breaking of the $SU(5)_{\rm ETC} \to SU(4)_{\rm ETC}$ 
and the running effect of the ETC gauge couplings at each breaking stage, 
and hence treated $\Lambda_2$ and $\Lambda_3$ as adjustable parameters.

Several analyses based on the ladder SD equation show 
that the critical binding strength for 
the MAC condensation for breaking ETC gauge symmetries is $k_{\crit} = 2\pi/3$~\cite{Gusynin:1982kp}, 
so that each $k^{(A,B)}_N$ for the MAC should be larger than $k^{\crit}$:
\beq
k^{(A,B)}_N > k^{\crit} = \frac{2\pi}{3}\,.
\eeq

Our assumption about the running of ETC gauge coupling is $k^{(5,1)}_5 \simeq k_{\crit}$ 
at $\Lambda_1 = 1000\TeV$ 
which corresponds to 
\beq
\alpha_{5(\rm ETC)}(\Lambda_1) = 0.436.
\label{ETC5critical}
\eeq 
There is of course possible error $1-20\%$~\cite{Appelquist:1988yc} 
of the estimation of $k^{\crit}$ due to the ladder approximation. 
Even if we take account of possible $30\%$ ambiguity of the critical coupling, 
$\alpha_{5(\rm ETC)}(\Lambda_1)$ could be lowered only to $\alpha_{5(\rm ETC)}(\Lambda_1) = 0.31$.
This is compared with the value $\alpha_{5(\rm ETC)}(\Lambda_1) = 0.1$ 
used in Ref.~\cite{Appelquist:1993sg,Appelquist:2003hn}.
   
Now we discuss the running effect of the ETC coupling on the criticality conditions,
starting with  $\alpha_{5(\rm ETC)}(\Lambda_1=1000\TeV) = 0.436$ as in Eq.(\ref{ETC5critical}).
The renormalization equation of each ETC gauge coupling are 
\beq
 &&
 \mu \frac{\partial}{\partial \mu} \alpha_{N(\rm ETC)} 
     = -b_N\,\alpha^2_{N(\rm ETC)}\,,\\
 &&
 b_N =\frac{1}{6\pi}
      \left[ 
       11 N - 2\left(\frac{1}{2} N^f_L + \frac{1}{2} N^f_R + \frac{N-2}{2} N^{\asymm}_R\right) 
      \right] (>0)\,,
\label{beta-N}
\eeq
where 
$N^f_{L(R)}$ is the number of left(right)-handed fermion 
with fundamental representation under $SU(N)_{\rm ETC}$ and
$N^{\asymm}_R$ is the number of right handed fermion 
with antisymmetric second rank representation under $SU(N)_{\rm ETC}$.
$(N^f_L,N^f_R,N^{\asymm}_R)$ for each ETC gauge group is 
\beq
(N^f_L,N^f_R,N^{\asymm}_R)
=\left\{\!\!\!
        \begin{array}{ll}
        \,(8,7,3)\, & {\rm for}\,\, N=5 \\[1pt]
         (8,10,2)   & {\rm for}\,\, N=4 \\[1pt]
        \,(8,9,1)\, & {\rm for}\,\, N=3 \\[1pt]
        \,(8,8,0)\, & {\rm for}\,\, N=2 
\end{array} \right.\,.
\eeq
$(N^f_L,N^f_R,N^{\asymm}_R)$ for $SU(2)_{\rm HC}$ gauge group 
with $N_\omega=2$ or $10$ is
\beq
\begin{array}{c}
N_\omega=2 \hspace*{3ex} N_\omega=10 \\[1ex]
{
(N^f_L,N^f_R,N^{\asymm}_R)
=\left\{\!\!\!
\begin{array}{lll}
         (0,12,0)& (0,20,0) & {\rm for}\,\,\, \Lambda_2 < \mu \\[1pt]
         (0,6,0) & (0,14,0) & {\rm for}\,\,\, \Lambda_3 < \mu < \Lambda_2 \\[1pt]
         (0,4,0) & (0,12,0) & {\rm for}\,\,\,\qquad\,\, \mu<\Lambda_3 
\end{array} \right.\,.
}
\end{array}
\eeq

In order to realize 
the desirable breaking $SU(4)_{\rm ETC} \to SU(3)_{\rm ETC}$ 
through the channel in  Eq.(\ref{4ETC-1}) with Eq.(\ref{43MAC})
at the scale $\Lambda_2$ lower than $\Lambda_1$,
we should take $SU(2)_{\rm HC}$ gauge coupling at $\Lambda_1$ as
$\alpha_{2(\rm HC)}(\Lambda_1)=0.59$ for $N_\omega = 2$ case 
and 
$\alpha_{2(\rm HC)}(\Lambda_1)=0.57$ for $N_\omega = 10$ case. 

Here we comment on $\Lambda_{2,3}$ in the case of $N_\omega = 2$ 
in Ref.~\cite{Appelquist:2003hn}.
In this case, $\Lambda_{2,3}$ is given as $\Lambda_2 = 850\,\TeV$ and $\Lambda_3 = 500\,\TeV$, 
so we obtain no  large hierarchy between $\Lambda_3$ and $\Lambda_{1,2}$ and hence no large mass difference 
between the third generation and the second/first generation.

In this paper, 
instead of $N_\omega = 2$, 
we take $N_\omega = 10$, 
which is 
the largest possible number of $\omega$ 
fulfilling $N_\omega \leq 10$ in order to keep the asymptotic freedom. 
Using settings of $N_\omega = 10$ 
and $k^{(A,B)}_{N}(\Lambda_i) \simeq k_{\crit}$, 
$\Lambda_{2,3}$ is given by
\footnote{
If we take 
account of possible $30\%$ ambiguity of $\kappa_{\rm crit}$
and $\Lambda_1 = 1000\,\TeV$, 
we obtain $\Lambda_2 = 360\,\TeV$ and $\Lambda_3 = 150\,\TeV$ 
for $N_\omega = 10$ case.
(This possible $30\%$ ambiguity of $k_{\rm crit}$ will produce 
$\Lambda_2 = 400\,\TeV$ and $\Lambda_3 = 220\,\TeV$ for $\Lambda_1 =1000\,\TeV$ 
in the case of $N_\omega = 2$.)
}
\beq
\Lambda_2 = 850\,\TeV\,,\,\Lambda_3 = 360\,\TeV\,.
\label{MAC-det-L2L3}
\eeq

After all the ETC gauge group breakings take place, we are left with
$SU(2)$ TC theory with $N_f = 8$  as discussed in Sec.\ref{SB-ETC}.
This TC has an intrinsic scale $\Lambda_{\rm TC}$ at two-loop level, 
which is taken as an effective cutoff in the walking/conformal TC~\cite{Appelquist:1996dq}, 
and we identify  $\Lambda_{\rm TC}$ with $\Lambda_3$. 
\beq
\Lambda_{\rm TC} = \Lambda_3 =  360\,\TeV\,:
\eeq

\begin{figure}[ht]
\begin{center}
\includegraphics[width=0.5\hsize,clip]{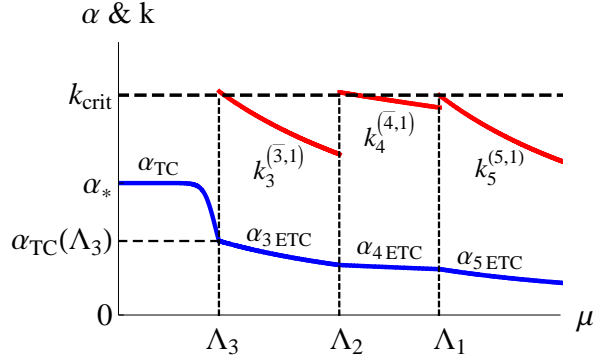}
\caption{Running behavior of both the ETC/TC gauge couplings 
         and the binding strengths for each stage  of  ETC breakings for the case of $N_\omega=10$
($\Lambda_1 = 1000\,\TeV$, $\Lambda_2 = 850\,\TeV$, and $\Lambda_3 = \Lambda_{\rm TC} = 360\,\TeV$.).
         $\alpha_* = 2\pi/5$ is the BZ-IRFP for $SU(2)$ gauge theory with $N_f =8$. 
         Each binding strength is
         $k^{(5,1)}_5$ in Eq.(\ref{5ETC-1}), 
         $k^{(\overline{4},1)}_4$ in Eq.(\ref{4ETC-1}) 
         and $k^{(\overline{3},1)}_3$ in Eq.(\ref{3ETC-1}) with the
         upper dashed line being the critical value $k_{\crit}=2\pi/3$.
         $\alpha_{\rm TC}(\Lambda_{\rm TC}) = \alpha_{\rm TC}(\Lambda_3) = 0.705$.
\label{ETC-MAC-run}}
\end{center}
\end{figure}

In the case of $N_\omega = 10$,  we have 
$\alpha_{N(\rm ETC)}\,,\,\alpha_{2(\rm HC)}$ 
at each scale of ETC breakings 
($\Lambda_1 =1000\,\TeV$, $\Lambda_2 = 850\,\TeV$ and $\Lambda_3 =360\,\TeV$) as 
\beq
\alpha_{5(\rm ETC)}(\Lambda_1) &=& \alpha_{4(\rm ETC)}(\Lambda_1) = 0.436\,,
\label{MAC-54-L1}\\
\alpha_{4(\rm ETC)}(\Lambda_2) &=& \alpha_{3(\rm ETC)}(\Lambda_2) = 0.476\,,
\label{MAC-43-L2}\\
\alpha_{3(\rm ETC)}(\Lambda_3) &=& \alpha_{(\rm TC)}(\Lambda_3) = 0.705\,,
\label{MAC-32-L3}\\
\alpha_{2(\rm HC)}(\Lambda_1) &=& 0.59\,,\\
\alpha_{2(\rm HC)}(\Lambda_2) &=& 0.596\,,\\
\alpha_{2(\rm HC)}(\Lambda_3) &=& 0.761\,.
\eeq
The running behavior of 
both the ETC/TC gauge couplings 
and the binding strengths for each  stage of  ETC breakings 
are shown in Fig.~\ref{ETC-MAC-run} for the case of $N_\omega=10$.

Now that all ETC breaking scales are fixed uniquely, we come to the discussion on
the third generation mass due to the TC condensate through ETC-induced four-fermion interactions: 
\beq 
{\cal L}^{\rm TC-3rd}_{\rm ETC}|_{\rm mass}
&=& 
    -G^{\rm ETC}_3
     \left[
     \Bigl(
      \overline{U}_R \gamma^\mu t_R
      +
      \overline{D}_R \gamma^\mu b_R
     \Bigr)
     \Bigl(
      \overline{q}_{3L} \gamma_\mu Q_L 
     \Bigr)     
     \right]
+ [{\rm h.c.}]\,,
\label{ETC-3rdmass-4F}
\eeq
where 
\beq
G^{\rm ETC}_3 = \frac{g^2_{3({\rm ETC})}}{2M^2_3} =\frac{1}{2\Lambda^2_3} 
              = \frac{4\pi\alpha_{3({\rm ETC})}(\Lambda_3)}{2M^2_3}\, ,  
\eeq
and $M_3=g_{3({\rm ETC})}\Lambda_3$ is the mass of the broken $SU(3)$ ETC gauge boson.
Now the $SU(2)$ TC with $N_f=8$ is a walking/conformal theory 
where the two-loop $\beta$-function possesses the BZ-IRFP $\alpha_* = 2\pi/5$ 
which is lower than the critical coupling evaluated in the ladder SD equation: 
$\alpha_* < \alpha^{\crit}=\pi/(3 C_2(F))$. 
However there is some ambiguity in evaluation of the critical coupling up to 1-20\% 
in the ladder approximation~\cite{Appelquist:1988yc}. 
In fact the critical coupling $\alpha^{\crit}$ decreases by 20\% 
when we define the critical coupling 
such that an anomalous dimension is $\gamma_m =1$ at two-loop level~\cite{Appelquist:1996dq}. 
So we may expect that the present TC triggers the technifermion condensate by its own dynamics.

After Fiertz rearrangement of  Eq.(\ref{ETC-3rdmass-4F}), 
we have the third generation quark mass given by 
\beq
m_{\rm 3rd} = \frac{4\pi\alpha_{3({\rm ETC})}(\Lambda_3)}{2M^2_3} \VEV{Q}_{M_3}\,,
\eeq
where $\VEV{Q}_{M_3}=\langle \overline{U}_R Q_L \rangle_{M_3}$
$=\langle \overline{D}_R Q_L \rangle_{M_3}$ is the technifermion condensate given by 
\beq
\VEV{Q}_{M_3}
&=&-\frac{N_{\rm TC}}{4\pi^2}
\int^{M^2_3}_{0} d x \frac{x \Sigma(x)}
{x+\Sigma(x)^2} \nonumber\\[1ex]
&=&
 -\frac{N_{\rm TC}}{4\pi^2}
 \left[
 \frac{2}{\gamma^{({\rm TC})}_m}
 \Bigl( \frac{M_3}{m_{\rm TC}} \Bigr)^{\gamma^{({\rm TC})}_m}
 +
 1 - \ln 2
 \right]
 m_{\rm TC}^3 
\,,
\label{TC-cond}
\eeq
for the dynamical mass function $\Sigma(x)$ of the technifermion parameterized as 
\beq
\Sigma(x) \sim  
\left\{\!\!\!
        \begin{array}{cl}
        m_{\rm TC} \Bigl(\dfrac{x}{m_{\rm TC}^2} \Bigr)^{\frac{1}{2}\gamma_m^{({\rm TC})}  - 1} 
         & {\rm for}\,\, x > m^2_{\rm TC} \\[3ex]
        \hspace*{-6ex}m_{\rm TC} & {\rm for}\,\, x < m^2_{\rm TC}
\end{array} \right.\,.
\eeq
with the anomalous dimension $\gamma_m^{({\rm TC})} \simeq 1$ for the walking/conformal $SU(2)_{\rm TC}$
technicolor.
Thus we have  $m_{\rm 3rd}$ 
\beq
m_{\rm 3rd} 
&\simeq& \frac{4\pi\alpha_{3({\rm ETC})}(\Lambda_3)}{2 M^2_3} \times \frac{m^2_{\rm TC} M_3}{\pi^2}
\nonumber\\
&\simeq& 0.1\,(\GeV) 
         \times \left( \frac{m_{\rm TC}}{500\,(\GeV)} \right)^2
         \times \left( \frac{360\,(\TeV)}{\Lambda_3} \right)\,,
\eeq
where we have used Eq.(\ref{MAC-32-L3}) and 
$M^2_3 = 4\pi\alpha_{3(\rm ETC)}(\Lambda_3) \times \Lambda^2_3$, with  a typical value of the technifermion mass  
in the one-family TC model ($N_{\rm TC} =2, N_f=8$) being  $m_{\rm TC} \simeq 500 {\rm GeV}$ which corresponds to 
$F_\pi^2 \simeq (246 {\rm GeV})^2/(N_f/2)$ (see, e.g., Eq.(\ref{TC-pi-decay-const})-(\ref{ev-fpiTC})).
Thus we can only have small mass for
the third generation  if  the TC  is the only origin of the EWSB in this type of ETC.

\section{The criticality of the EWSB}
\label{crit-EWSB}
Let us now discuss the EWSB in the top-mode ETC 
where both quarks and techniquarks having the same topcolor $SU(3)_1$ are put in 
the same representation of ETC.

As to the topcolor breaking, we assume that 
$\vev{\Phi} = \Lambda_{\rm C} \neq 0\,,\,(\Lambda_3 > \Lambda_{\rm C})$ trigger the topcolor breaking as 
\beq
SU(3)_1 \!\!\!\!\!\!\!\!\!\!\!&\times&\!\!\!\!\!\!\!\!\!\!\! SU(3)_2 \,,
\nonumber \\
&\big\downarrow&\!\!\!\!\!\!\!\!\!\!\!\!\Lambda_{\rm C} \\
&SU(3)_{\rm QCD}\!& \,\,,
\nonumber
\eeq
where 
$SU(3)_1$ is stronger than $SU(3)_2$ and 
the gauge coupling of $SU(3)_1 \times SU(3)_2$ are given by $h_1$ and $h_2$, respectively.
This topcolor breaking generates 
the 8 massive gauge bosons ( colorons ) 
and 8 massless gauge bosons ( gluons ).
The colorons mass $M_{\rm C}$ is given by 
\beq
M_{\rm C} = \sqrt{h^2_1 + h^2_2}\,\Lambda_{\rm C}\,,
\eeq
and the $SU(3)_{\rm QCD}$ gauge coupling is given by
\beq
g_{\rm QCD} = h_1 \sin \theta = h_2 \cos \theta \,,
\eeq
where we defined the mixing angle $\theta$ as 
\beq
\cot \theta \equiv \frac{h_1}{h_2} ( > 1)\,.
\eeq  

After all the ETC and topcolor breakings occurred 
( A mechanism of ETC breaking is shown in Sec.~\ref{crit-sETC}), 
we obtain 
the $SU(2)_{\rm TC} \times SU(3)_{\rm QCD} \times SU(2)_L \times U(1)_Y$ invariant four fermion interaction :
\beq
{\cal L}^{4f} = \sum_{i,j}{\cal L}^{i-j}_{\rm ETC} + \sum_{i,j}{\cal L}^{i-j}_{\rm topC}\,,
\label{all4f-TMETC}
\eeq
where 
$i,j$ is TC or SM and 
for example ${\cal L}^{\rm TC-SM}_{\rm ETC}$ represents four fermion interactions 
between technifermions and SM-fermions via massive ETC gauge bosons. For the moment we shall concentrate on the 
four-fermion interactions which lead to the diagonal mass of the quarks/leptons in Eq.(\ref{all4f-TMETC}): 
\beq 
{\cal L}^{\rm TC-SM}_{\rm ETC}|_{\rm mass}
&=& 
    -G^{\rm ETC}_1
    \left[
     \Bigl(
      \overline{U}_R \gamma^\mu u_R
      +
      \overline{D}_R \gamma^\mu d_R
     \Bigr)
     \Bigl(
      \overline{q}_{1L} \gamma_\mu Q_L 
     \Bigr)  
     +
     \Bigl(
      \overline{t}_R \gamma^\mu u_R
     \Bigr)
     \Bigl(
      \overline{q}_{1L} \gamma_\mu q_{3L}
     \Bigr)
     \right]
\nonumber\\
&& 
    -G^{\rm ETC}_2
     \left[
     \Bigl(
      \overline{U}_R \gamma^\mu c_R
      +
      \overline{D}_R \gamma^\mu s_R
     \Bigr)
     \Bigl(
      \overline{q}_{2L} \gamma_\mu Q_L 
     \Bigr)
     +
     \Bigl(
      \overline{t}_R \gamma^\mu c_R
     \Bigr)
     \Bigl(
      \overline{q}_{2L} \gamma_\mu {q}_{3L}
     \Bigr)      
     \right]
\nonumber\\ 
&&
    -G^{\rm ETC}_3
     \left[
     \Bigl(
      \overline{U}_R \gamma^\mu t_R
      +
      \overline{D}_R \gamma^\mu b_R
     \Bigr)
     \Bigl(
      \overline{q}_{3L} \gamma_\mu Q_L 
     \Bigr)
     +
     \frac{1}{3}
     \Bigl(
      \overline{t}_R \gamma^\mu t_R
     \Bigr)
     \Bigl(
      \overline{q}_{3L} \gamma_\mu {q}_{3L}
     \Bigr)     
     \right]
\nonumber\\
&&  -\sum_i
    G^{\rm ETC}_i
     \Bigl(
      \overline{E}_R \gamma^\mu e_{iR}
     \Bigr)
     \Bigl(
      \overline{l}_{iL} \gamma_\mu L_L 
     \Bigr)  + [{\rm h.c.}]\,,
\label{ETC-mass-4F}
\eeq
where
\beq
G^{\rm ETC}_i= \frac{c_i \times 4\pi\alpha_{N ({\rm ETC})}}{2M^2_i} 
\,,\left(i=1,2,3; c_1=2\,,\,c_2=\frac{3}{2}\,,\,c_3=1\right)\,,
\label{ETC-Gi}
\eeq
$g_{N(\rm ETC)}$ is the $SU(N)_{\rm ETC}$ gauge coupling at $\Lambda_i$. 
From Sec.~\ref{crit-sETC}, 
our hierarchy reads $\Lambda_1\gg \Lambda_2\gg \Lambda_3$. 
The sideway ETC gauge bosons mass is given by
\beq 
M_i = g_{N(\rm ETC)} \Lambda_i\,. 
\eeq

Let us  discuss the criticality of the top quark condensate. 
We concentrate on $(\overline{f} f)^2$-type four fermion interaction part in Eq.(\ref{all4f-TMETC}):
\beq
{\cal L}^{4f}|_{\rm self}
&=&
\sum_{i=1,2,3}
\left[
\frac{1}{6-i} G^{\rm ETC}_i 
\Bigl((\overline{q}_{iL} u_{iR})^2 + (\overline{q}_{iL} d_{iR})^2 
      + (\overline{l}_{iL} e_{iR})^2
\Bigr) 
\right.
\nonumber\\
&&\hspace*{20ex}
\left.
+ G_{\rm topC} \Bigl((\overline{q}_{iL} u_{iR})^2 + (\overline{q}_{iL} d_{iR})^2 \Bigr)
\right]
\,,
\label{TMETCcond-4f}
\eeq
where 
\beq
G_{\rm topC}
= \frac{h^2_1 \cos^2 \theta}{4M^2_{\rm C}}
= \frac{4\pi\alpha_{\rm QCD}}{4M^2_{\rm C}} \cot^2 \theta 
= \frac{4\pi}{4M^2_{\rm C}} \cdot \kappa_3\,,
\label{G-topC}
\eeq
and
\beq
\kappa_3 
\equiv \alpha_{\rm QCD} \cot^2 \theta
 =     \alpha_{SU(3)_1} \cos^2 \theta 
 =     \alpha_{SU(3)_2} \cos^2 \theta \cot^2 \theta 
\,,
\label{def-kappa3}
\eeq
and
$\alpha_{a} \equiv g^2_{a}/4\pi\,(a= {\rm QCD}\,,\,SU(3)_1\,,\,SU(3)_2)$.

Now, we consider the gap equation for the four fermion interaction Eq.(\ref{TMETCcond-4f}):
\beq
1
&=&
\left[ \frac{1}{6-i}G^{\rm ETC}_i + G_{\rm topC} \right]
\times 
\frac{2 N_c M^2_{\rm C}}{4\pi^2} 
\Bigl[ 1 - \frac{m^2_{\rm dyn}}{M^2_{\rm C}} \ln \frac{M^2_{\rm C}}{m^2_{\rm dyn}}\Bigr]\,,
\,\,(\text{for quarks})\,,
\\
1
&=&
\left[ \frac{1}{6-i}G^{\rm ETC}_i  \hspace*{8.2ex}\right] 
\times \hspace*{2ex}
\frac{2 M^2_{\rm C}}{4\pi^2} \hspace*{1ex}
\Bigl[ 1 - \frac{m^2_{\rm dyn}}{M^2_{\rm C}} \ln \frac{M^2_{\rm C}}{m^2_{\rm dyn}}\Bigr]\,,
\,\,(\text{for leptons})\,,
\label{gap-TMETC}
\eeq
where $N_c (=3) $ is the number of colors, 
and $m_{\rm dyn}$ is the dynamical mass of each fermion.
We define the dimensionless four fermion coupling $g_{fi}\,,$ 
($fi$ stands for the SM fermions) as
\beq
g_{ui} 
 &\equiv& 
   \left[ G^{\rm ETC}_i + G_{\rm topC} \right] \frac{2 N_c M^2_{\rm C}}{4\pi^2} 
 = N_c \cdot g^{\rm ETC}_i \,\,+ \hspace*{0.625ex} N_c \cdot \hspace*{0.625ex} \frac{\kappa_3}{2 \pi} \,,
\label{dg-uiTMETC}\\
g_{di} 
 &\equiv& 
   \left[ G^{\rm ETC}_i + G_{\rm topC} \right] \frac{2 N_c M^2_{\rm C}}{4\pi^2} 
 = N_c \cdot g^{\rm ETC}_i \,\,+ \hspace*{0.625ex} N_c \cdot \hspace*{0.625ex} \frac{\kappa_3}{2 \pi} \,,
\label{dg-diTMETC}\\
g_{ei} 
 &\equiv&
   \left[ G^{\rm ETC}_i \hspace*{8ex} \right] \,\,\,\,\frac{2 M^2_{\rm C}}{4\pi^2}\,\,\,\, 
 = \hspace*{4ex} g^{\rm ETC}_i \,,
\label{dg-eiTMETC}
\eeq 
where 
\beq
g^{\rm ETC}_i 
= G^{\rm ETC}_i \cdot \frac{2 M^2_{\rm C}}{4\pi^2} 
= \frac{c_i \alpha_{N ({\rm ETC})}}{\pi} \cdot \left( \frac{M_{\rm C}}{M_i} \right)^2\,.
\eeq

We can realize the situation that the top quark is the only SM fermion to condense, 
only if  the following conditions are met:
\beq
g_t > g^{\crit}_t\,\,\,,\,\,\,\,\,
g_{b,\tau} &<& g^{\crit}\,\,,
\label{cond-3TMETC}\\
g_{\tilde{f}} &<& g^{\crit} \,,\,({\tilde{f}}=u,d,c,s,e,\mu)\,.
\label{cond-iTMETC}
\eeq
As we discussed in Sec.~\ref{crit-sETC} 
we have the hierarchical ETC breaking scale, 
so that we have  a hierarchy of 
$g^{\rm ETC}_i$ 
\beq
g^{\rm ETC}_3 
&>& g^{\rm ETC}_2 
    = g^{\rm ETC}_3 \times \frac{3}{4} 
                    \cdot \frac{c_2}{c_3} \cdot \left( \frac{M_3}{M_2} \right)^2\nonumber\\
&>& g^{\rm ETC}_1 
    = g^{\rm ETC}_3 \times \frac{3}{5}
                    \cdot \frac{c_1}{c_3} \cdot \left( \frac{M_3}{M_1} \right)^2\,,
\label{hierarchy-gETC}
\eeq 
i.e., the condensate of the third generation quarks/leptons  is favored to that of others. 
Then we concentrate on Eq.(\ref{cond-3TMETC}). 

In the NJL case $g^{\crit}$ is $g^{\crit} = 1$ in Eq.(\ref{cond-3TMETC}), 
however, dynamics in the present case is the gauged NJL model ( SM gauge + NJL ). 
We recall the critical coupling (critical line) in the gauged NJL model~\cite{Kondo:1988qd}:
\beq
g_f^{\crit} 
= \frac{1}{4}\left(1+\sqrt{1-\frac{\alpha_f}{\pi/3}}\right)^2\,,
\label{crit-gNJL}
\eeq 
where 
\beq
\alpha_{f=t,c,u}(\mu)&=&\frac{4}{3}\,\alpha_{\rm QCD}(\mu) +\, \frac{1}{9}\, \alpha_Y(\mu) \,,
\label{uimac}\\
\alpha_{f=b,s,d}(\mu)&=&\frac{4}{3}\,\alpha_{\rm QCD}(\mu) - \frac{1}{18} \alpha_Y(\mu)\,,
\label{dimac}\\
\alpha_{f=e,\mu,\tau}(\mu)&=&\hspace*{11ex} +\,\,\, \frac{1}{2}\,\, \alpha_Y(\mu)\,,
\label{eimac}
\eeq 
In order to obtain the top quark condensation, 
$\kappa_3$ should satisfy Eq.(\ref{cond-3TMETC}) which read: 
\beq
 \kappa_3 + 2 \pi \cdot g^{\rm ETC}_3
 &>& \frac{2 \pi}{N_c} \cdot g_t^{\crit}\,,
\label{topmac-TMETC}\\
 \kappa_3 + 2 \pi \cdot g^{\rm ETC}_3
 &<& \frac{2 \pi}{N_c} \cdot g_b^{\crit}\,,
\label{botmac-TMETC}\\
 g^{\rm ETC}_3
 &<& \hspace*{5ex} g_{\tau}^{\crit}\,,
\label{taumac-TMETC}
\eeq

In the present case 
\beq 
g^{\rm ETC}_3 
\simeq 5.0 \times 10^{-6} 
           \times \left( \frac{M_{\rm C}}{5\,(\TeV)} \right)^2 
           \times \left( \frac{360\,(\TeV)}{\Lambda_3} \right)^2\,,
\label{tb-4ETC}
\eeq 
so we can neglect $g^{\rm ETC}_3$ in Eq.(\ref{topmac-TMETC}), (\ref{botmac-TMETC}) and (\ref{taumac-TMETC}). 
The coloron mass is constrained by the experiment $M_{\rm C} /\cot \theta > 837\,\GeV$~\cite{Bertram:1998wf}, 
which implies  $M_{\rm C} \gtrsim 3\,\TeV$ in our case (see Eq.(\ref{alpha3bc})). 
We shall take 
\beq
M_{\rm C} =5 \,\TeV.
\label{coloronmass}
\eeq
In order to trigger the top quark condensation in the present model, 
$\kappa_3$ must satisfy 
\beq
 \kappa_3 
 > \frac{2 \pi}{N_c} \cdot g_t^{\crit}
\quad {\rm and} \quad
 \kappa_3 
 < \frac{2 \pi}{N_c} \cdot g_b^{\crit}\,,
\label{kappa3-TMETC}
\eeq
where in the case of $M_{\rm C} \simeq 5\,\TeV$,
\beq
 g_t^{\crit}(M_{\rm C}) \simeq  0.942 
\quad , \quad
 g_b^{\crit}(M_{\rm C}) \simeq  0.943 \,, 
\label{gcrit-tbTMETC}
\eeq
which correspond to 
\beq
 \alpha_{t}(M_{\rm C}) \simeq  0.119 
\quad , \quad  
 \alpha_{b}(M_{\rm C}) \simeq  0.117 \,, 
\label{alpha_tbfu}
\eeq
where we have used inputs: 
$\alpha_Y(M_Z) = 0.0101684 \pm 0.0000014, \alpha_{\rm QCD}(M_Z) = 0.1176 \pm 0.0020$~\cite{Yao:2006px}.
Eq.(\ref{kappa3-TMETC}) and (\ref{gcrit-tbTMETC}) show 
a constraint on $\kappa_3$ as
\beq
1.973 < \kappa_3 < 1.975\,.
\label{onlytopmac-TMETC}
\eeq
From Eq.(\ref{def-kappa3}), 
this result shows 
\beq
22.22<\cot^2 \theta < 22.25\,,
\quad \text{and} \quad
2.062<\alpha_{SU(3)_1}(M_{\rm C}) < 2.064\,.
\label{alpha3bc}
\eeq 

Now we discuss the criticality of the technifermion condensate.
In this model, technifermions have topcolor charge as shown in Table.~\ref{particle-content}, 
so that
the technifermion condensate is triggered at the scale $\mu$ 
if $\alpha_{U/D/E/N} (\mu) > \pi/3$ is satisfied, 
where $\alpha_{U/D/E/N} (\mu)$ is given by  
\beq
\alpha_U (\mu) 
&=& 
\frac{3}{4}\,\alpha_{\rm TC}(\mu) + \frac{4}{3}\,\alpha_{SU(3)_1}(\mu) \, + \,\frac{1}{9} \, \alpha_Y(\mu)
\,\\
\alpha_D (\mu) 
&=& 
\frac{3}{4}\,\alpha_{\rm TC}(\mu) + \frac{4}{3}\,\alpha_{SU(3)_1}(\mu) - \frac{1}{18}\, \alpha_Y(\mu)
\,,\\
\alpha_E (\mu) 
&=& 
\frac{3}{4}\,\alpha_{\rm TC}(\mu) \hspace*{15ex} \,+ \,\frac{1}{2} \, \alpha_Y(\mu)
\,,\\
\alpha_N (\mu) 
&=& 
\frac{3}{4}\,\alpha_{\rm TC}(\mu)\,.
\eeq

From Eq.(\ref{alpha3bc}) and Eq.(\ref{MAC-32-L3}) 
we estimate $\alpha_{SU(3)_1} (\mu > M_{\rm C})$ and $\alpha_{\rm TC} (\mu < \Lambda_3)$, respectively.
The $U(1)_Y$ contribution is negligible. 
Then we find that the combined coupling of TC and $SU(3)_1$ 
is rather strong already at $\mu =\Lambda_3$: $\alpha_U(\Lambda_3) \simeq 0.972 \sim \pi/3$. 
In fact we find $\alpha_{U/D} (\mu) > \pi/3$ at $\mu \simeq 80\,\TeV$ 
which implies that the dynamical mass of the techniquark 
and hence the decay constant $f_{\pi_T}$ is 
on this order $f_{\pi_T} \sim m_{\rm TC} \simeq 80\,\TeV$, 
which is extremely large compared with the weak scale and  
is a disaster.

Therefore, the framework in this section cannot give us a desirable result.
In order to avoid this problem, 
we should change the topcolor charge assignment of the technifermions, 
so that the technicolor criticality could be unaffected 
by the strong topcolor which is required to be near criticality for triggering the top quark condensate. 
In the Section~\ref{fullU-TC2}, 
we consider such a new TC2 model, 
although an explicit ETC model building is not attempted in this paper.

\section{Twisted flavor-universal TC2}
\label{fullU-TC2}

As discussed in Sec.~\ref{crit-EWSB}, 
we should change the topcolor charge assignment of the techniquarks.
In this section, 
as the first step to build  an explicit ETC model having such a topcolor assignment,
we here consider a new type TC2 model, twisted flavor-universal TC2 model,  
in order to forbid a techniquark condensate at too large scale.
In addition to the topcolor $SU(3)_1\times SU(3)_2$ we here introduce an extended hypercharge sector 
 $U(1)_{Y1}\times U(1)_{Y2}$~\cite{Braam:2007pm} to be spontaneously broken into the SM hypercharge
symmetry $U(1)_Y$, in such a way that SM fermions  carry the flavor-universal
$SU(3)_1 \times U(1)_{Y1}$, while technifermions do the opposite charges $SU(3)_2 \times U(1)_{Y2}$. 
The charge assignments in this twisted flavor-universal TC2 model
are shown in Table.~\ref{particle-content-FUTC2}.
\begin{table}
\begin{center}
\begin{tabular}{| c | c | c | c | c | c | c |
}
\hline
field & $ SU(2)_{\rm TC} $ & $ SU(3)_1$  & $SU(3)_2$ & $SU(2)_L$ & $U(1)_{Y1}$ 
& $U(1)_{Y2}$ 
\\
\hline 
$Q_L$ & 2 & 1 & 3 & 2 & 0 & $1/6$   
\\
$U_R$ & 2 & 1 & 3 & 1 & 0 & $2/3$  
\\
$D_R$ & 2 & 1 & 3 & 1 & 0 & $-1/3$  
\\ 
\hline
$L_L$ & 2 & 1 & 1 & 1 & 0 & $-1/2$
\\
$E_R$ & 2 & 1 & 1 & 1 & 0 & $-1$
\\
$N_R$ & 2 & 1 & 1 & 1 & 0 & 0
\\
\hline
\, & \, & \, & \, & \, & \, & \, 
\\[-2.5ex] 
\hline
$q_{iL}$ & 1 & 3 & 1 & 2 & $1/6$ & 0 
\\
$u_{iR}$ & 1 & 3 & 1 & 1 & $2/3$ & 0 
\\
$d_{iR}$ & 1 & 3 & 1 & 1 & $-1/3$ & 0 
\\ 
\hline
$l_{iL}$ & 1 & 1 & 1 & 2 & $-1/2$ & 0 
\\
$e_{iR}$ & 1 & 1 & 1 & 1 & $-1$ & 0 
\\
\hline
\end{tabular}
\caption{Particle contents in the twisted flavor-universal TC2 model.
         \label{particle-content-FUTC2}}
\end{center}
\end{table}%

Since the techniquarks and quarks have different topcolor charges, 
it is highly nontrivial to put them into a single representation of the ETC. 
We would need larger picture to unify them. 
For the moment 
we shall not try such an explicit model building 
but discuss possible consequences 
if the ETC type interactions communicate the SM fermions and the technifermions. 
Such a possibility may also be realized 
by the composite model for the SM fermions and the technifermions~\cite{Yamawaki:1982tg}. 
All the setting of symmetry breakings is made analogously to 
the ETC breakings of the model studied in the Sec.~\ref{A-TMETC} and Sec.~\ref{crit-sETC}. 
Such an approach is the same as the conventional TC2 model building.

The symmetry breaking is assumed as follows : 
\beq
{\rm ETC}_1 \times  SU(3)_1 \times SU(3)_2 \!\!\!\!\!
&\times &\!\!\!\!\! SU(2)_L \times U(1)_{Y1} \times U(1)_{Y2} 
\nonumber\\[1ex]
&\big\downarrow&\!\!\!\Lambda_1 \nonumber \\
{\rm ETC}_2 \times  SU(3)_1 \times SU(3)_2 \!\!\!\!\!
&\times &\!\!\!\!\! SU(2)_L \times U(1)_{Y1} \times U(1)_{Y2} 
\nonumber\\[1ex]
&\big\downarrow&\!\!\!\Lambda_2 \label{FUTC2-bp}\\
{\rm ETC}_3 \times  SU(3)_1 \times SU(3)_2 \!\!\!\!\!
&\times &\!\!\!\!\! SU(2)_L \times U(1)_{Y1} \times U(1)_{Y2} 
\nonumber\\[1ex]
&\big\downarrow&\!\!\!\Lambda_3 \nonumber\\
SU(2)_{\rm TC} \times  SU(3)_1 \times SU(3)_2 \!\!\!\!\!
&\times &\!\!\!\!\! SU(2)_L \times U(1)_{Y1} \times U(1)_{Y2} \,,
\nonumber \\
&\big\downarrow&\!\!\!\Lambda_{\rm C} < \Lambda_3 \nonumber\\
SU(2)_{\rm TC} \times \hspace*{4ex} SU(3)_{\rm QCD} \,\hspace*{1ex}
&\times&\!\!\!\!\! SU(2)_L \times \hspace*{3ex} U(1)_Y\,,
\nonumber
\eeq
where we do not specify each ETC gauge group and dynamical mechanism of the ETC breakings in this section.
Also, we do not discuss the dynamical mechanism of the topcolor/extended hypercharge breakings.
Each gauge coupling of topcolor/extended hypercharge is follows: 
$SU(3)_1 ( SU(3)_2 )$ gauge coupling is $h_1\,( h_2 )$ and 
$U(1)_{Y1} ( U(1)_{Y2} )$ gauge coupling is $g'_{Y1}\,( g'_{Y2} )$, 
where $SU(3)_1/U(1)_{Y1}$ is stronger than $SU(3)_2/U(1)_{Y2}$.
The colorons and $Z'$ mass $M_{\rm C},M_{Z'}$ are 
\beq
M_{\rm C}    &=& \sqrt{h^2_1 + h^2_2}\,\Lambda_{\rm C}\,,\\
M_{Z'} &=& \sqrt{{g'}^2_{Y1} + {g'}^2_{Y2}}\,\Lambda_{\rm C}
\eeq
and the $SU(3)_{\rm QCD}$ and $U(1)_Y$ gauge couplings are given by
\beq
g_{\rm QCD} = \!\!\!&h_1 \sin \theta& \!\!\!= h_2 \cos \theta \,, \\
g'_Y        = \!\!\!&g'_{Y1} \sin \eta&\!\!\! = g'_{Y2} \cos \eta \,\,
\eeq
where we defined the mixing angles $\theta$ and $\eta$ as 
\beq
\cot \theta \equiv \frac{h_1}{h_2} ( >1 )
\quad , \quad
\cot \eta \equiv \frac{g'_{Y1}}{g'_{Y2}} ( >1 )\,.
\eeq  
As to the ${\rm ETC}_3 \to SU(2)_{\rm TC}$ breaking, the sideway gauge boson mass $M_3$ 
is a free parameter at this moment and   
we assume here $M_3 \gtrsim M_{\rm C} = M_{Z'}$ for simplicity.

At $ \Lambda_{\rm C}$ the topcolor/extended hypercharge gauge groups 
as well as the ETC group spontaneously break down, 
so that we have  $(\overline{f} f)^2$-type four fermion interactions: 
\beq
{\cal L}^{4f}
&=&
\sum_{i=1,2,3}
\Bigl[
G^{\rm ETC}_i 
\Bigl((\overline{q}_{iL} u_{iR})^2 + (\overline{q}_{iL} d_{iR})^2 
      + (\overline{l}_{iL} e_{iR})^2
\Bigr) 
+
G^s_{\rm topC} \Bigl((\overline{q}_{iL} u_{iR})^2 + (\overline{q}_{iL} d_{iR})^2 \Bigr)
\Bigr.
\nonumber\\
&&\hspace*{7ex}+
\Bigl.
G^{su}_{Z'} (\overline{q}_{iL} u_{iR})^2 
+ 
G^{sd}_{Z'} (\overline{q}_{iL} d_{iR})^2
+
G^{sl}_{Z'} (\overline{l}_{iL} e_{iR})^2 
\Bigr]\,,
\label{FUTC2-4f}
\eeq
where 
$q_{iL}=(u_i,d_i)^T_L\,,\,l_{iL}=(\nu_i,e_i)_{iL}\,,\,u_3=t\,,\,d_3=b\,,\,\nu_3=\nu_\tau\,,\,e_3=\tau\,,\cdots$.  
$G^{\rm ETC}_i = 4\pi\alpha_{{\rm ETC}_i}(\Lambda_i)/(2M^2_i) = 1/(2\Lambda^2_i)\,(i=1,2,3)$ are  
the four-fermion couplings of the SM fermions 
generated by the ETC breaking at each scale $\Lambda_i$, 
$G^s_{\rm topC}$ is the four-fermion coupling generated 
by the topcolor $SU(3)_1 \times SU(3)_2$ breaking at $\Lambda_{\rm C}$ 
and 
$G^{sf}_{\rm Z'}$ is the four-fermion coupling 
of the fermions: $f\,,\,(f = t,b,c,s,u,d,\cdots)$ generated 
by the extended hypercharge gauge symmetry $U(1)_{Y1} \times U(1)_{Y2}$ breaking at $\Lambda_{\rm C}$. 
 
We rewrite 
$G^{\rm ETC}_i$, $G^s_{\rm topC}$ and $G^s_{Z'}$ 
by $g^{\rm ETC}_i$, $\kappa_3$ and $\kappa_1$, respectively
\beq
G^{\rm ETC}_i 
= \frac{2\pi^2}{M^2_{\rm C}} \cdot g^{\rm ETC}_i
\,\,\,,\,\,\,
G^s_{\rm topC} 
= \frac{4\pi}{4 M^2_{\rm C}} \cdot \kappa_3\,
\,\,\,,\,\,\,
G^{sf}_{Z'} 
= \frac{4\pi}{2 M^2_{Z'}}\cdot \kappa_1 Y^f_LY^f_R \,,        
\eeq
where 
\beq
\kappa_3 
\equiv \alpha_{\rm QCD} \cot^2 \theta 
 \!\!\!&=&\!\!\!\alpha_{SU(3)_1} \cos^2 \theta
 =     \alpha_{SU(3)_2} \cos^2 \theta \cot^2 \theta\,,\\
\kappa_1 \equiv \,\,\, \alpha_Y \cot^2 \eta
 &=&\alpha_{Y1} \cos^2 \eta \,\,\,\,\,
 =     \alpha_{Y2} \cos^2 \eta \cot^2 \eta\,, 
\eeq
and 
$\alpha_{m} \equiv g^2_{m}/4\pi\,(m= {\rm QCD}\,,\,SU(3)_1\,,\,SU(3)_2)$ 
and 
$\alpha_{n} \equiv g'^2_{n}/4\pi\,(n= Y\,,\,Y1\,,\,Y2)$.
We define the dimensionless four-fermion coupling $g_{fi}$ as
\beq
g_{u,c,t} 
 &\equiv& 
   \left[ N_c G^{\rm ETC}_i + N_c G^s_{\rm topC} + G^{su}_{Z'} \right] \frac{2M^2_{\rm C}}{4\pi^2} 
 = N_c \cdot g^{\rm ETC}_i + N_c  \cdot \frac{\kappa_3}{2 \pi} 
    + \, \frac{1}{9}\, \cdot \frac{\kappa_1}{\pi}\,,
\label{dg-ui}\\
g_{d,s,b} 
 &\equiv& 
   \left[N_c G^{\rm ETC}_i + N_c G^s_{\rm topC} + G^{sd}_{Z'} \right] \frac{2M^2_{\rm C}}{4\pi^2} 
 = N_c \cdot g^{\rm ETC}_i + N_c \cdot \frac{\kappa_3}{2 \pi}
   - \frac{1}{18} \cdot \frac{\kappa_1}{\pi}\,,
\label{dg-di}\\
g_{e,\mu,\tau} 
 &\equiv&
   \left[ \hspace*{2.5ex}G^{\rm ETC}_i \hspace*{11ex} + G^{sl}_{Z'} \right] \frac{2M^2_{\rm C}}{4\pi^2} 
 = \hspace*{4ex} g^{\rm ETC}_i \hspace*{10.5ex}
   + \, \frac{1}{2}\, \cdot \frac{\kappa_1}{\pi}\,,
\label{dg-ei}
\eeq
In terms of (\ref{dg-ui})--(\ref{dg-ei}) 
we have the conditions 
that top quark  is the only SM fermion to condense, 
\beq
g_t > g_t ^{\rm crit} \,\,\,,\,\,\,\,\,
g_{b,\tau} &<& g^{\crit}_{b,\tau}\,\,,
\label{cond-3FUTC2}\\
g_{u,d,c,s,e,\mu} &<& g^{\crit}_{u,d,c,s,e,\mu}\,, 
\label{cond-iFUTC2}
\eeq
where the critical lines for SM fermions, 
$g^{\crit}_{u,c,t}$, 
are given in Eq.(\ref{crit-gNJL})--(\ref{eimac}), 
which yield
\beq
 g^{\crit}_{u,c,t}(M_{\rm C})      & \simeq & 0.942\,, \nonumber\\
 g^{\crit}_{d,s,b}(M_{\rm C})      & \simeq & 0.943\,, \nonumber\\
 g^{\crit}_{e,\mu,\tau}(M_{\rm C}) & \simeq & 0.997\,, 
\label{mac-MB}
\eeq
for the values of 
$\alpha_{u,c,t} \simeq 0.120\,,\,\alpha_{d,s,b} \simeq 0.118\,,\,\alpha_{e,\mu,\tau} \simeq 0.005$ 
(see Eq.(\ref{alpha_tbfu})). 

The breaking in Eq.(\ref{FUTC2-bp}) generates the hierarchical four-fermion couplings, 
\beq
G^{\rm ETC}_1 < G^{\rm ETC}_2 < G^{\rm ETC}_3 \,,
\label{ETC4fermi-ls}
\eeq
which implies that the condensation of the third generation fermions are favored to other generations.
If we can realize large hierarchical ETC breaking $\Lambda_{1,2}/\Lambda_3 \gg 1$, 
the condition Eq.(\ref{cond-iFUTC2}) can easily fulfilled. 
Then we concentrate on the condition Eq.(\ref{cond-3FUTC2}) 
in order to consider the criticality of the top quark condensation.

\begin{figure}[h]
\begin{center}
\includegraphics[width=0.4\hsize,clip]{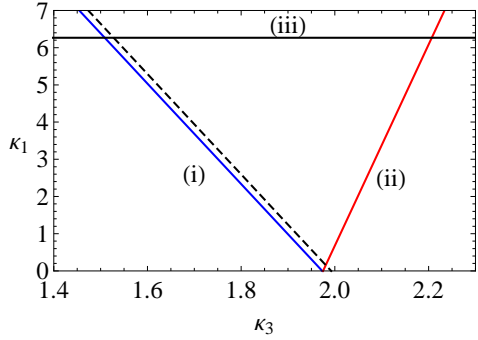}
\caption{The gap triangle in twisted flavor-universal TC2 model ($M_{\rm C} = M_{Z'} = 5\,\TeV$). 
         The region above (i) represents $\VEV{t} \neq 0$, 
         the region below (ii) represents $\VEV{b} \neq 0$ and 
         the region above (iii) represents $\VEV{\tau} \neq 0$.
         Only the top quark forms a condensation in 
         the area (``gap triangle'') enclosed by the lines (i), (ii) and (iii). 
         The dashed line stands for  the solution line of  the gauged NJL model~\cite{Nonoyama:1989dq} 
         for a finite dynamical mass of the top quark  
         $m_t \simeq 170\,\GeV$ and $M_{\rm C} = 5\,\TeV$ in Eq.(\ref{crit-line-nonzero}).
\label{gap-triangle}}
\end{center}
\end{figure}

Eq.(\ref{cond-3FUTC2}) are explicitly written as 
\beq
 2 \pi \cdot g^{\rm ETC}_3 
 \,+\, \kappa_3 \,+ \frac{1}{N_c} \cdot \frac{2}{9} \cdot \kappa_1 
 &>& \frac{2 \pi}{N_c} \cdot g^{\crit}_t\,,
\label{topmac-FUTC2}\\
 2 \pi \cdot g^{\rm ETC}_3 
 \,+\, \kappa_3\, - \frac{1}{N_c} \cdot \frac{1}{9} \cdot \kappa_1 
 &<& \frac{2 \pi}{N_c} \cdot g^{\crit}_b\,,
\label{botmac-FUTC2}\\
 g^{\rm ETC}_3 
 \hspace*{6.5ex}+
 \hspace*{4.5ex}
 \frac{1}{2} \cdot \kappa_1 
 &<& \hspace*{1ex} \pi \hspace*{0.5ex}\cdot \hspace*{0.5ex}g^{\crit}_\tau\,,
\label{taumac-FUTC2}
\eeq
where $g^{\crit}_{t,b,\tau}$ are given by Eq.(\ref{mac-MB}) and
the value of $g^{\rm ETC}_3 = G^{\rm ETC}_3 \cdot {2 M^2_{\rm C}}/{4\pi}
= {1}/({4\pi^2}) \cdot \left({M_{\rm C}}/{\Lambda_3}\right)^2$ is estimated as
\beq
g^{\rm ETC}_3 
\simeq 3.5 \times 10^{-6} 
           \times \left( \frac{M_{\rm C}}{5\,(\TeV)} \right)^2 
           \times \left( \frac{360\,(\TeV)}{\Lambda_3} \right)^2\,,
\label{gETC3-TWTC}
\eeq
which is negligibly small for $\Lambda_3 > 360\,\TeV$. 
Then the parameter space area $(\kappa_3\,,\,\kappa_1)$ constrained by  
Eq.(\ref{topmac-FUTC2}), (\ref{botmac-FUTC2}) and (\ref{taumac-FUTC2}) 
is represented by the triangle (``gap triangle'') in Fig.~\ref{gap-triangle}
in the case of $\Lambda_{\rm C} = 1\,\TeV$ for $M_{\rm C} = 5\,\TeV$(See Eq.(\ref{coloronmass})).
The dashed line in Fig.~\ref{gap-triangle} stands for 
the top quark mass coming from top condensate 
${\hat m}_t$ ($\simeq m_t^{\rm exp}  = 172\pm2.5 \GeV)$) determined by
the SD gap equation:~\cite{Nonoyama:1989dq}
\beq
g^{\crit}_{{\hat m}_t} 
= \frac{1}{4}
  \frac{(1+\sqrt{1-\alpha_t/(\pi/3)})^2 - (1-\sqrt{1-\alpha_t/(\pi/3)})^2 (P+Q)}
       {1-P+\frac{3-\sqrt{1-\alpha_t/(\pi/3)}}{1+\sqrt{1-\alpha_t/(\pi/3)}}Q}\,,
\label{crit-line-nonzero}
\eeq 
where $P,Q$ are given by
\beq
&&P \equiv 
\frac{
\Gamma(1-\sqrt{1-\alpha_t/(\pi/3)})\Gamma(3/2+\frac{1}{2}\sqrt{1-\alpha_t/(\pi/3)})^2
}{
\Gamma(1+\sqrt{1-\alpha_t/(\pi/3)})\Gamma(3/2-\frac{1}{2}\sqrt{1-\alpha_t/(\pi/3)})^2
}
\Bigl(\frac{{\hat m}^2_t}{M^2_{\rm C}}\Bigr)^{\sqrt{1-\alpha_t/(\pi/3)}}\, ,\nonumber\\
&&
Q \equiv
\frac{(1+\sqrt{1-\alpha_t/(\pi/3)})^2}{4(1-\sqrt{1-\alpha_t/(\pi/3)})} \frac{{\hat m}^2_t}{M^2_{\rm C}}\,.
\eeq

Let us consider the criticality of the technifermion condensate in the present model.
First, we consider the criticality in the $\Lambda_{\rm C} < \mu < \Lambda_3 = \Lambda_{\rm TC}$, 
where the ETC breaking scale $\Lambda_3$ is identified with the intrinsic scale of the $\Lambda_{\rm TC}$ 
defined by the two-loop beta function~\cite{Appelquist:1996dq}.  
In this region 
ETC breaks down to TC while topcolor/extended hypercharge gauge symmetry does not.
Since $g^{\rm ETC}_3$ is small as seen Eq.(\ref{gETC3-TWTC}), 
relevant is the gauge dynamics described by $\alpha_{U,D,E,N} (\mu)$: 
\beq
\alpha_U (\mu) 
&=& 
\frac{3}{4}\,\alpha_{\rm TC}(\mu) + \frac{4}{3}\,\alpha_{SU(3)_2}(\mu) \, + \,\frac{1}{9} \, \alpha_{Y2}(\mu)
\,\\
\alpha_D (\mu) 
&=& 
\frac{3}{4}\,\alpha_{\rm TC}(\mu) + \frac{4}{3}\,\alpha_{SU(3)_2}(\mu) - \frac{1}{18}\, \alpha_{Y2}(\mu)
\,,\\
\alpha_E (\mu) 
&=& 
\frac{3}{4}\,\alpha_{\rm TC}(\mu) \hspace*{15ex} \,+ \,\frac{1}{2} \, \alpha_{Y2}(\mu)
\,,\\
\alpha_N (\mu) 
&=& 
\frac{3}{4}\,\alpha_{\rm TC}(\mu)\,.
\eeq 
Now we observe that   
$\alpha_{\rm TC} (\mu) < \alpha_{\rm TC}(\Lambda_{\rm C}) = 1.18$ 
which is estimated by two-loop running of the $\alpha_{\rm TC}(\mu)$ 
with the boundary condition $\alpha_{\rm TC} (\Lambda_{3}) 
= \alpha_{\rm TC}(\Lambda_{\rm TC}) \simeq 0.782 \times \alpha_* \simeq 0.983$%
~\cite{Appelquist:1996dq}
where $\alpha_*$ is the BZ-IRFP of the one-family $SU(2)$ TC as the walking/conformal TC,
$\alpha_{SU(3)_2} (\mu) < \alpha_{SU(3)_2} (\Lambda_{\rm C}) = 0.092 - 0.094$  corresponding to 
the allowed value of $1.6 < \kappa_3 < 2.2$, 
and $\alpha_{Y2} (\mu)= {\cal O}(10^{-2})$. 
Therefore $\alpha_{U,D,E,N}(\mu)$ do not exceed 
the critical coupling for the region $\Lambda_{\rm C} < \mu < \Lambda_3$: 
\beq
\alpha_{U,D,E,N}(\mu) <
\alpha_U (\Lambda_{\rm C}) \!\!\!
&\simeq&\!\!\! \frac{3}{4} \alpha_{\rm TC}(\Lambda_{\rm C}) 
+ \frac{4}{3} \alpha_{SU(3)_2}(\Lambda_{\rm C}) 
+ \frac{1}{9} \alpha_{Y2}(\Lambda_{\rm C}) \nonumber\\
&\simeq& 1.007 - 1.010
\,\,\,
< \,\,\, \frac{\pi}{3} \simeq 1.047\,.
\eeq
and hence the TC + weak topcolor gauge interactions 
in the region of $\Lambda_{\rm C} < \mu < \Lambda_3$ 
does not trigger the technifermion condensate.
 
Next, we consider the criticality in the $\mu < \Lambda_{\rm C}$, 
where ETC/topcolor/extended hypercharge gauge symmetry break down.
The four-fermion interactions involving only the technifermions at $\Lambda_{\rm C}$
are given by
\beq
{\cal L}^{4F}
&=&
G^{\rm ETC}_{\rm TC} 
\Bigl((\overline{Q}_L U_R)^2 + (\overline{Q}_L D_R)^2 
      + (\overline{L}_L E_R)^2 + (\overline{L}_L N_R)^2 
\Bigr)
\nonumber\\
&&\hspace*{6ex}
+
G^w_{\rm topC} \Bigl((\overline{Q}_L U_R)^2 + (\overline{Q}_L D_R)^2 \Bigr) 
\nonumber\\
&&\hspace*{10ex}+
G^{wU}_{Z'} (\overline{Q}_L U_R)^2 
+ 
G^{wD}_{Z'} (\overline{Q}_L D_R)^2
+
G^{wE}_{Z'} (\overline{L}_L E_R)^2 \,,
\label{FUTC2-4fTC}
\eeq
where $(F = U,D,E,N)$
\beq
G^{\rm ETC}_{\rm TC} 
= \frac{2\pi^2}{M^2_{\rm C}} \cdot g^{\rm ETC}_{\rm TC}
\,\,\,,\,\,\,
G^w_{\rm topC}
= \frac{4\pi}{4M^2_{\rm C}} \cdot \frac{\alpha^2_{\rm QCD}}{\kappa_3}
\,\,\,,\,\,\,
G^{wF}_{Z'} 
= \frac{4\pi}{2M^2_{Z'}} \cdot \frac{\alpha^2_Y}{\kappa_1} Y^F_LY^F_R \,.       
\eeq
Now, the dimensionless four-fermion coupling of technifermions $U,D,E$ are defined as
\beq
g_U  
 &=& N_{\rm TC} N_c \cdot g^{\rm ETC}_{\rm TC} 
   + \frac{N_{\rm TC} N_c}{2} \cdot \frac{\alpha^2_{\rm QCD}}{\pi \kappa_3} 
   + \,\frac{N_{\rm TC}}{9}\, \cdot \frac{\alpha^2_Y}{\pi \kappa_1}\,,
\label{dg-U}\\
g_D 
 &=& N_{\rm TC} N_c \cdot g^{\rm ETC}_{\rm TC}  
   + \frac{N_{\rm TC} N_c}{2} \cdot \frac{\alpha^2_{\rm QCD}}{\pi \kappa_3} 
   - \frac{N_{\rm TC}}{18} \cdot \frac{\alpha^2_Y}{\pi \kappa_1}\,,
\label{dg-D}\\
g_E 
 &=& N_{\rm TC} \hspace*{2.5ex} \cdot g^{\rm ETC}_{\rm TC} \hspace*{18ex}
   +\,\frac{N_{\rm TC}}{2}\, \cdot \frac{\alpha^2_Y}{\pi \kappa_1}\,,
\label{dg-E}
\eeq 
where $g^{\rm ETC}_{\rm TC} \simeq g^{\rm ETC}_3$ up to ${\cal O}(1)$ coefficient 
and Eq.(\ref{gETC3-TWTC}) with $\Lambda_3 = 360\,\TeV$ 
shows $N_c g^{\rm ETC}_3 \simeq 10^{-5}$. 
The allowed region in Fig.~\ref{gap-triangle} shows 
$N_c \alpha^2_{\rm QCD}/(\pi \kappa_3) \simeq 10^{-2}$, 
$\alpha^2_Y/(\pi \kappa_1) \simeq 10^{-2}$. 
The four-fermion couplings are negligible compared with the gauge dynamics, TC + SM gauge interactions,
whose running effects yield $\alpha_{U,D}(\mu) > \pi/3$ at $\mu \simeq 470\,\GeV$ 
and hence the dynamical masses of techniquarks 
\beq
m_{\rm TC} 
\simeq {\cal O}(470\,\GeV)\,,
\label{TQcond-scale}
\eeq
which reproduces the weak scale.

Now we explicitly show that the model realizes TC2 scenario.
In this model, the TC theory is walking below $\Lambda_3 = \Lambda_{\rm TC}$.
This fact shows that the TC theory develops non-zero anomalous dimension $\gamma^{(\rm TC)}_m$.  
Using $\gamma^{(\rm TC)}_m$, 
the technipion ($\pi_T$) decay constant: $f_{\pi_T}$ 
(by Pagels-Stokar formula~\cite{Pagels:1979hd}) 
and technifermion condensation: 
$\VEV{F}_{M_3}\,,\,(F = U,D,E,N)$ 
are represented as
\beq
 f^2_{\pi_T}
 &=&
  \frac{N}{4\pi^2}\int^{\infty}_{0} \!\!\!dx 
  x \frac{\Sigma^2(x) - \frac{x}{4}\frac{d}{dx}\Sigma^2(x)}{(x+\Sigma^2(x))^2}
 \nonumber\\[1ex]
 &=&  
  \frac{N}{8\pi}
  \left[
  \frac{3-\frac{\gamma^{(\rm TC)}_m}{2}}{(3-\gamma^{(\rm TC)}_m)^2}
  \frac{1}{\sin(\frac{\pi}{3-\gamma^{(\rm TC)}_m})}
  +
  \frac{2}{\pi}\left(\ln 2 - \frac{1}{2}\right)
  \right]
  m^2_{\rm TC}
 \,,
\label{TC-pi-decay-const}
\\[1ex]
\VEV{F}_{M_3}
&=&-\frac{N}{4\pi^2}
\int^{M^2_3}_{0}\!\!\! d x \frac{x \Sigma(x)}
{x+\Sigma(x)^2} \nonumber\\[1ex]
&=&
 -\frac{N}{4\pi^2}
 \left[
 \frac{2}{\gamma^{(\rm TC)}_m}  
 \Bigl( \frac{M_3}{m_{\rm TC}} \Bigr)^{\gamma^{(\rm TC)}_m} 
 +
 1 - \ln 2
 \right]
 m_{\rm TC}^3 
\,,
\label{TC-cond2}
\eeq
with $m_{\rm TC}$ being a dynamical mass of the technifermion $\Sigma(x =m_{\rm TC}^2) =m_{\rm TC}$, 
where 
\beq
\Sigma(x) \sim  
\left\{\!\!\!
        \begin{array}{cl}
        m_{\rm TC} \Bigl(\dfrac{x}{m_{\rm TC}^2} \Bigr)^{\frac{1}{2}\gamma^{(\rm TC)}_m - 1} 
         & {\rm for}\,\, x > m^2_{\rm TC} \\[3ex]
        \hspace*{-6ex}m_{\rm TC} & {\rm for}\,\, x < m^2_{\rm TC}
\end{array} \right.\,.
\label{dmass1}
\eeq
Eq.(\ref{TQcond-scale}) and Eq.(\ref{TC-pi-decay-const}) give $f_{\pi_T}$ as 
\beq 
f_{\pi_T} \simeq 115\,\GeV \qquad \text{for} \qquad m_{\rm TC} \simeq 470\,\GeV\,,
\label{ev-fpiTC}
\eeq
where we have used that 
the anomalous dimension of TC: $\gamma^{(\rm TC)}_m$ 
is $\gamma^{(\rm TC)}_m \simeq 1$ 
because 
the combined gauge coupling of TC with weak topcolor $SU(3)_2$ 
for $\Lambda_{\rm C} < \mu < \Lambda_3$ 
and 
the combined one of TC with $SU(3)_{\rm QCD}$ for $\mu < \Lambda_{\rm C}$ 
are near critical coupling.
Eq.(\ref{ev-fpiTC}) implies that 
in order to reproduce the weak scale $(246\,\GeV)^2 = 4f^2_{\pi_T} + f^2_{\pi_t}$, 
the top-pion ($\pi_{t}$) decay constant $f_{\pi_t}$ should be 
\beq 
f_{\pi_t} \simeq 87\,\GeV\,,
\label{tpi-TWTC2}
\eeq 
and hence the top quark mass ${\hat m}_t$ 
coming from the top quark condensation should be 
\beq
{\hat m}_t \simeq 167\,\GeV\,,
\eeq
where we have used the Pagels-Stokar formula 
\beq
f^2_{\pi_t} = \frac{3}{8\pi^2} {\hat m}_t^2 \ln \frac{M^2_{\rm C}}{{\hat m}_t^2}\,,
\label{top-cond}
\eeq
with $M_{\rm C} = 5\,\TeV$ 
for the constant top quark mass function induced by NJL-type topcolor dynamics 
corresponding to $\gamma_m \simeq 2$%
~\footnote{Including QCD $\log$ correction  
decreases $f_{\pi_t}$  by 10\% from 
$f_{\pi_t}$ in Eq.(\ref{top-cond}).}.

Let us discuss the third generation quark masses.
The ETC induced four-fermion interactions responsible for the third generation quark masses 
are given by ${\cal L}_{\rm 3rd-mass}$:
\beq 
{\cal L}_{\rm 3rd-mass}
&=&  G^{\rm ETC}_3
      \left[
        (\overline{U}_R  Q_L) (\overline{q}_{3L} t_R)
        +
        (\overline{D}_R  Q_L) (\overline{q}_{3L} b_R)
      \right] \nonumber\\[0.5ex]
&&   +(G^s_{\rm topC} + G^s_{Z'}) 
      (\overline{t}_R  q_{3L}) (\overline{q}_{3L} t_R)
     + [{\rm h.c.}] 
\label{mass-4f}
\eeq
as discussed in Sec.~\ref{crit-sETC}. 
The bottom quark may acquire mass by technifermion condensation 
through Eq.(\ref{mass-4f}) in the same way as the ordinary ETC model.
\beq
m_b = G^{\rm ETC}_3 \langle \overline{D}_R Q_L \rangle
\eeq
On the other hand,
the top quark may acquire mass 
by both technifermion condensation through Eq.~(\ref{mass-4f}) and top quark condensation, 
\beq
m_t &=& G^{\rm ETC}_3 
      \langle \overline{U}_R Q_L \rangle 
      +
      (G^{\rm ETC}_3 + G^s_{\rm topC} + G^s_{Z'})
      \langle \overline{t}_R q_L \rangle \nonumber\\
    &\equiv& 
      \hspace*{5ex} m^{(0)}_t \hspace*{4.2ex} + \hspace*{10ex} {\hat m}_t
\,,
\eeq
under the conditions Eq.(\ref{topmac-FUTC2})--(\ref{taumac-FUTC2}).

Eq.(\ref{mass-4f}) gives 
the bottom quark mass $m_b(\Lambda_3)$ as 
\beq
m_b (\Lambda_3) 
    &=& G^{\rm ETC}_3 \langle \overline{D}_R Q_L \rangle_{M_3} 
    \simeq \frac{4\pi}{2M^2_3} \cdot (0.782 \times \alpha_*) \cdot 
      \langle \overline{D}_R Q_L \rangle_{M_3} \nonumber\\
    &\simeq& 0.11\,\GeV 
    \times \left( \frac{m_{\rm TC}}{470\,(\GeV)} \right)^2
    \times \left( \frac{360\,(\TeV)}{\Lambda_3} \right)\,,
\label{ETC-bottom}
\eeq
and ETC induced top quark mass ${\hat m}^{(0)}_t(\Lambda_3)$ 
is the same as $m_b(\Lambda_3)$.

However, 
due to the topcolor dynamics of NJL-type, 
these masses are greatly amplified as follows.
In the NJL-type dynamics there arises large anomalous dimension $\gamma_m \simeq 2$ 
near the critical coupling even in the symmetric phase~\cite{Kikukawa:1989fw} 
and 
the bare mass at the scale $\Lambda$ is greatly enhanced at lower energy scale $\mu ( \ll \Lambda )$ 
\beq
m_\mu = Z^{-1}_m m_\Lambda\,
\quad , \quad
Z^{-1}_m ( \Lambda/\mu )
= \left( \frac{\Lambda}{\mu} \right)^{2} 
\cdot 
\left[\frac{\ln (\Lambda/\Lambda_{\rm QCD})}{\ln (\mu/\Lambda_{\rm QCD})}\right]^{-A/2}\,,
\eeq
where the logarithmic correction to $\gamma_m = 2$ comes from QCD correction 
with $A=24/(33-2N_f) >1$ and 
$A =8/7>1$ for $N_f=6$ ($\mu < m_{\rm TC}$) 
and $A =24/13>1$ for $N_f=10$ ($m_{\rm TC} < \mu < M_{\rm C}$).
Note that the gauged NJL model with $A > 1$ is renormalizable~\cite{Kondo:1992sq,Kondo:1991yk}.
In the case at hand $\Lambda = M_{\rm C}$ and $\mu = m_t$ 
and hence $Z^{-1}_m (M_{\rm C}/m_t)  
\simeq 560$, 
so we have 
\beq
m_b(m_t) = {\hat m}^{(0)}_t(m_t) \simeq 5\,\GeV\,,
\eeq 
for $\Lambda_3 \simeq 4500\,\TeV$ and $M_{\rm C} = 5\,\TeV$. 

If we arrange the ETC breaking scales $\Lambda_{1,2}$ somewhat higher than 
that of the third generation $\Lambda_3$
so that the combined four-fermion interactions of topcolor, extra $U(1)$ and ETC
are off the criticality, 
then the mass of the second and the first generation fermions 
would have no large enhancement due to anomalous dimension 
and hence give reasonable hierarchy compared with the top and bottom.

Leptons would acquire masses from the technilepton condensate $\VEV{E}$  
whose gauge coupling 
$\alpha_E \simeq 3/4 \times \alpha_{\rm TC} \simeq 3/4 \times \alpha_* = 3\pi/10$ 
is smaller than the critical coupling $\pi/3$ in the ladder approximation. 
However, as discussed in Sec.~\ref{crit-sETC} 
there is some ambiguity in evaluation of the critical coupling up to 1-20\% 
in the ladder approximation~\cite{Appelquist:1988yc}. 
In fact the critical coupling $\alpha^{\crit}$ decreases by 20\% 
when we define the critical coupling 
such that an anomalous dimension is $\gamma_m =1$ at two-loop level~\cite{Appelquist:1996dq}. 
Thus, up to this ambiguity 
leptons may acquire mass somewhat smaller than the mass of quarks~\cite{Appelquist:1997fp}. 
\section{Top-pion Mass}
\label{top-pion mass}
We here discuss a novel effect of the large anomalous dimension of the topcolor 
on the estimation of the top-pion mass $m_{\pi_t} < 70\,\GeV$, 
which is extremely smaller than the conventional estimation $\simeq 200-300\,\GeV$~\cite{Hill:2002ap}. 
This applies to generic TC2 model not restricted to ours.

The top-pion appears in the generic TC2 model\cite{Hill:1994hp,Hill:1991at}, 
since both the top quark condensate and technifermion condensate break respective global symmetries, 
which results in two kinds of three  Nambu-Goldstone (NG) bosons. 
Three of them (mainly boundstates of technifermions) are absorbed into $W,Z$ bosons as usual.  
The rest three are pseudo NG bosons (mainly boundstates of top and bottom), 
which is called the top-pion.
In general, 
the former three NG bosons decay constant $f_{\pi_T}$  
and 
the top-pion decay constant $f_{\pi_t}$   
reproduce the weak scale $(246\,\GeV)^2 = N_D f^2_{\pi_T} + f^2_{\pi_t}$, 
where $N_D$ is the number of the technifermion EW-doublet. 

The top-pion mass may be estimated by the Dashen formula~\cite{Dashen:1969eg} 
\beq
m^2_{\pi_t} f^2_{\pi_t} = - m^{(0)}_t \VEV{t}\,,
\label{Dashen-formula-TC2}
\eeq
where $m^{(0)}_t$ is the ETC-induced top quark mass acting as the bare mass 
for the NJL-type topcolor dynamics.

Now, we recall that 
the renormalization of generic mass parameter $m$ is carried out with keeping such a relation as 
\beq
m_\Lambda (\overline{\psi} \psi)_\Lambda 
= m_\mu (\overline{\psi} \psi)_\mu
\quad,\qquad
\Lambda \gg \mu\,,
\label{reno-mass}
\eeq
where $\psi$ is the generic fermion and 
the suffix $\Lambda (\mu)$ represents a bare (renormalized) quantity 
at $\Lambda (\mu)$ ($\mu$ is the reference renormalization point). 
Eq.(\ref{reno-mass}) shows that 
when we write the renormalization of the mass parameter $m$ as 
\beq
m_\Lambda = Z_m m_\mu\,,
\label{reno-mass-TC2}
\eeq
then the renormalization of the composite operator $(\overline{\psi} \psi)$ 
should be given by 
\beq
(\overline{\psi} \psi)_\Lambda 
= 
Z^{-1}_m (\overline{\psi} \psi)_\mu\,, 
\label{reno-cond-TC2}
\eeq
where $Z^{-1}_m = (\Lambda/\mu)^{\gamma_m}$ 
is the renormalization factor 
and $\gamma_m$ is the anomalous dimension of the mass parameter.
In the TC2 model, 
the condensation $\VEV{t}$ at $M_{\rm C}$ is represented by~\cite{Hill:1994hp}
\beq
\VEV{t}|_{M_{\rm C}} 
=-\frac{N_c}{4\pi^2}
\int^{M^2_{\rm C}}_{0}\!\!\! d x \frac{x \Sigma(x)}
{x+\Sigma(x)^2} 
= -\frac{N_c}{4\pi^2} \cdot {\hat m}_t M^2_{\rm C}\,,
\eeq
for the constant top quark mass ${\hat m}_t (= m_t - m^{(0)}_t)$
induced by the NJL-type topcolor dynamics corresponding $\gamma_m \simeq 2$. 
Eq.(\ref{reno-cond-TC2}) relates $\VEV{t}|_{M_{\rm C}}$ to $\VEV{t}|_{{\hat m}_t}$ as
\beq
\VEV{t}|_{M_{\rm C}} 
= Z^{-1}_m \VEV{t}|_{{\hat m}_t}
= \left( \frac{M_{\rm C}}{{\hat m}_t} \right)^2 \VEV{t}|_{{\hat m}_t}\,,
\eeq
and the condensation $\VEV{t}|_{{\hat m}_t}$ is represented by  
\beq
\VEV{t}|_{m_t} 
=-\frac{N_c}{4\pi^2}
\int^{{\hat m}^2_t}_{0}\!\!\! d x \frac{x \Sigma(x)}
{x+\Sigma(x)^2} 
= -\frac{N_c}{4\pi^2} \cdot {\hat m}^3_t \,.
\eeq
On the other hand, 
Eq.(\ref{reno-mass-TC2}) relates the ETC-induced top quark mass 
${\hat m}^{(0)}_t (M_{\rm C})$ to ${\hat m}^{(0)}_t ({\hat m}_t)$ as
\beq
m^{(0)}_t(M_{\rm C}) 
= Z_m m^{(0)}_t({\hat m}_t) 
= \left( \frac{{\hat m}_t}{M_{\rm C}} \right)^2 m^{(0)}_t({\hat m}_t)\,.
\eeq 
Therefore, 
we observe that 
the right-hand side of Eq.(\ref{Dashen-formula-TC2}) 
is {\it renormalization point independent}: 
\beq
m^{(0)}_t (M_{\rm C}) \cdot \VEV{t}|_{M_{\rm C}} 
= 
m^{(0)}_t ({\hat m}_t) \cdot \VEV{t}|_{{\hat m}_t}\,,
\eeq
which shows that in the TC2 model we have the generic form as 
\beq
m^2_{\pi_t} f^2_{\pi_t} = m^{(0)}_t({\hat m}_t) \cdot \frac{N_c}{4\pi^2} {\hat m}^3_t\,. 
\label{Dashen-formula-TC2prime}
\eeq

Now, the top-pion decay constant $f_{\pi_t}$ is evaluated in exactly the same way 
as in the original top quark condensate paper~\cite{Miransky:1988xi}.
The Pagels-Stokar Formula gives $f_{\pi_t}$ 
as a function of ${\hat m}_t$ (top quark mass coming from the top quark condensate) and $M_{\rm C}$ :
\beq
f^2_{\pi_t}
=  
\frac{3}{8\pi^2} {\hat m}^2_t \ln \frac{M^2_{\rm C}}{{\hat m}^2_t}\,,
\label{PS-formula-TC2}
\eeq
for the constant top quark mass induced by the NJL-type topcolor dynamics corresponding $\gamma_m \simeq 2$. 

Thus
Eq.(\ref{Dashen-formula-TC2prime}) and (\ref{PS-formula-TC2}) 
give us the generic form of the top-pion mass as  
\beq
m^2_{\pi_t} 
&=& 
2 \cdot m^{(0)}_t({\hat m}_t) \cdot \frac{{\hat m}^3_t}{{\hat m}^2_t \cdot \ln (M^2_{\rm C}/{\hat m}^2_t)} 
=
\frac{ m^{(0)}_t({\hat m}_t) \cdot {\hat m}_t}{\ln (M_{\rm C}/{\hat m}_t)}\,.
\label{gen-mpit}
\eeq
Experimentally, 
a model-independent lower limit~\cite{Bertram:1998wf}  of the  coloron mass $M_{\rm C}$ is
\beq
M_{\rm C}/\cot \theta  &>& 837\,\GeV\, ({\rm flavor-universal})\,,
\nonumber
\\
 M_{\rm C}/\cot \theta
 & >& 450\,\GeV \,({\rm flavor-non-universal})\,,
\label{C-mass}
\eeq
for the flavor-universal coloron and the flavor-non-universal coloron, respectively,
where $\cot \theta = h_1/h_2 > 1$,
because $h_{1(2)}$ represents the gauge coupling of the strong (weak) topcolor $SU(3)_{1(2)}$.

Hence Eq.(\ref{gen-mpit}) and (\ref{C-mass}) yield 
an upper bound of top-pion mass 
\beq
m^2_{\pi_t} 
< \frac{ m^{(0)}_t({\hat m}_t) \cdot {\hat m}_t}{\ln (M_{\rm C}/{\hat m}_t)}
= \frac{ m^{(0)}_t({\hat m}_t) \cdot \left[ m_t - m^{(0)}_t({\hat m}_t) \right]}{\ln (M_{\rm C}/{\hat m}_t)}
\,, 
\label{FU-pit} 
\eeq 
where physical top quark mass is $m_t = {\hat m}_t + m^{(0)}_t= 172\,\GeV$.
Eq.(\ref{FU-pit}) has the maximum value at $m^{(0)}_t({\hat m}_t) \simeq 70\,\GeV$ as 
\beq
m^2_{\pi_t} 
< \frac{m^{(0)}_t({\hat m}_t) \cdot {\hat m}_t}{\ln (M_{\rm C}/{\hat m}_t)} 
&\simeq& (60\,\GeV)^2\, ,
\nonumber\\
&\simeq& (70\,\GeV)^2\, ,
\label{toppi-allTC2}
\eeq
for the flavor-universal coloron case and
the flavor non-universal case, respectively.
Eq.(\ref{toppi-allTC2}) is a very conservative upper bound 
universal to generic model of TC2 not restricted to specific TC2 model, 
since in the generic TC2 model we have actually $\cot \theta = h_1/h_2 > 4$ 
instead of $\cot \theta = h_1/h_2 > 1$ in order to trigger the top quark condensate 
(see Eq.(\ref{alpha3bc}))~\cite{Hill:2002ap}.

Our estimate above is quite different from the conventional estimate 
which is made at the scale $M_{\rm C}$: $m^{(0)}_t(M_{\rm C}) \cdot \VEV{t}|_{M_{\rm C}}$, 
where $\VEV{t}|_{M_{\rm C}} = [N_c ({\hat m}_t)^3/(4\pi^2)]\cdot (M_{\rm C} /{\hat m}_t)^2 $
up to logarithm. 
$\VEV{t}|_{M_{\rm C}}$ is larger than that estimated at ${\hat m}_t$
by $Z^{-1}_m \simeq (M_{\rm C} /{\hat m}_t)^2 $ as it should. 
The crucial point is the evaluation of $m^{(0)}_t(M_{\rm C})$, 
which is usually made independently of $m^{(0)}_t({\hat m}_t)$
and taken as on the order of $m_b\simeq 5 \,\GeV$.
However, if we do this, 
the physical mass of the bottom $m_b({\hat m}_t)$ and the ETC-driven
mass of the top $m^{(0)}_t({\hat m}_t)$ should be enhanced 
by the same factor $Z^{-1}_m \simeq (M_{\rm C}/{\hat m}_t)^2$ into absurdly large value.
Or,  $m^{(0)}_t(M_{\rm C})$ must be taken to be $Z_m$ times smaller than 
the physical value $m^{(0)}_t({\hat m}_t)$ 
which should be a small portion of the top quark mass in the TC2 scenario 
$m^{(0)}_t({\hat m}_t) \ll m_t \simeq 172\,\GeV$ . 
This is the effect of the large anomalous dimension. 
Anyway the result should be the same as ours as far as we correctly take account of 
{\it renormalization-point independence} 
of the operator $m_t^{(0)}  \overline{t} t$
which is multiplicatively renormalized, 
since the gauged NJL model in this case is renormalizable (See the previous footnote \ref{foot-RP}).

\section{Summary and discussions}
\label{sum}

We have experimented a straightforward explicit ETC model building 
which incorporates the top quark condensate via universal coloron type topcolor $SU(3)_1 \times SU(3)_2$
which is spontaneously broken to the ordinary $SU(3)_{\rm QCD}$. 
All the quarks and techniquarks were assigned to have only
$SU(3)_1$ which is much stronger than $SU(3)_2$ to trigger the top quark condensate. 

The criticality conditions of MAC for the $SU(5)$ ETC and $SU(2)$ hypercolor dynamics 
realized the successive ETC breakings down to $SU(2)$ TC 
which is walking/conformal near the conformal window. 
Imposing the criticality conditions at each step of ETC breaking  
predicted the ETC breaking scales somewhat larger than those of Ref.~\cite{Appelquist:1993sg,Appelquist:2003hn} 
and hence very small ETC-driven masses of the third generation quarks/leptons, 
of order ${\cal O}(10^{-1}\,\GeV)$, 
in spite of the enhancement of large anomalous dimension $\gamma_m \simeq 1$ of the walking/conformal TC.

Imposing the topcolor $SU(3)_1$ to be near the criticality, 
we realized the top quark condensate so as to give a realistic mass to the top quark.
However, the techniquarks then feel both the strong topcolor near criticality
as well as  the equally strong walking/conformal TC near the criticality. 
Such combined strong gauge interactions trigger 
the techniquark condensate at the scale ridiculously large compared with the weak scale.

Then we considered an alternative model of TC2, 
with coloron and $Z^\prime$ of flavor-universal type, 
where the quarks have strong $SU(3)_1 \times U(1)_{Y_1}$ interactions, 
while techniquarks do weak $SU(3)_2 \times U(1)_{Y_2}$ interactions, 
with both spontaneously broken to the SM gauge theories  $SU(3)_{\rm QCD}\times U(1)_{Y}$.
Since explicit ETC model of this charge assignment is rather involved, 
we only considered here the effective theory of TC2 assuming that 
similar ETC breaking can take place in a larger picture of a certain ETC gauge group.
In such a framework we discuss the mass of third generation quarks 
can be realistic even when all the ETC breaking scales are 
somewhat larger than usually considered as we demonstrated in the explicit ETC model. 
A key observation was that
the ETC-driven mass of quarks are regarded as the bare mass of the topcolor sector, 
which then can be enormously enhanced 
by the large anomalous dimension $\gamma_m \simeq 2$ of the NJL-type dynamics 
of the broken topcolor, 
if the effective four-fermion coupling is near the criticality: 
$m^{(0)}_t (m_t) = Z_m^{-1} m^{(0)}_t(M_{\rm C}) $ 
with $Z^{-1}_m \simeq (M_{\rm C}/m_t)^{\gamma_m} \simeq (M_{\rm C}/m_t)^2$, 
which is typically  $Z^{-1}_m \simeq 500$ for the coloron
mass $M_{\rm C} > 4 \,\TeV$.
We then obtained realistic masses 
$m_b \simeq 5\,\GeV$ as well as $m_t \simeq 172\,\GeV$ 
whose main part $\simeq 167\,\GeV$ comes from the top quark condensate 
and the rest $\simeq 5\,\GeV (\simeq m_b)$ is 
the ETC origin mass enhanced by the anomalous dimension $\gamma_m \simeq 2$. 
If we arrange the ETC breaking scales somewhat higher than 
that of the third generation 
so that the combined four-fermion interactions of topcolor, extra $U(1)$ and ETC
are off the criticality, 
then the mass of the second and the first generation fermions 
would have no large enhancement due to anomalous dimension 
and hence give reasonable hierarchy compared with the top and bottom.

Another possibility to modify the top-mode ETC type model 
would be to put the fourth generation in stead of the technifermion 
and the SM fermions into the same representation of a horizontal group, say $SU(4)$, 
in such a way that the fourth quarks have the same strong flavor-universal topcolor $SU(3)_1$ 
as that of the three generations quarks. 
Then the fourth generation quark condensate triggered by the topcolor 
would play the role of the technifermion condensate. 
In order that only the fourth quark ($t^\prime,\, b^\prime$) and the top quark should condense, 
we should arrange the ETC-type interactions to discriminate them from others 
in such a way that the effective four-fermion couplings are arranged as
$g_{t^\prime}>g^{\rm crit}_{t^\prime}, g_{b^\prime}> g^{\rm crit}_{b^\prime}, g_t > g^{\rm crit}_t$, 
where $g_i^{\rm crit}$ ($i=t^\prime, b^\prime, t$)
is the critical line of the gauged NJL model having the SM gauge interaction contributions (see Sec. 4). 
Of course, the fourth generation neutrino should have Majorana condensate 
in order to avoid the light fourth neutrino. 
Explicit ETC-type model having successive symmetry breaking of the horizontal symmetry would be interesting.

Finally, 
we found a novel effect of the large anomalous dimension $\gamma_m \simeq 2$ 
of the NJL-type dynamics on the evaluation of the top-pion mass 
through the Dashen formula together with the Pagels-Stokar formula. 
The Dashen formula contains the bare mass (ETC-induced mass) times top quark condensate. 
Usual estimate of this combination is made 
at the scale of the topcolor breaking (coloron mass) $M_{\rm C}$: 
The condensate has an enhancement of quadratic divergence of NJL-type $\sim M_{\rm C}^2$, 
while the bare mass  at $M_{\rm C}$ scale 
was just assumed to be the order of the physical bottom mass $\sim m_b$, 
which is, however, 
enormously enhanced as much as $10^2-10^3$ times by 
the renormalization effect $Z_m^{-1} \simeq (M_{\rm C}/m_t)^2$ 
due to the same quadratic divergence when evaluated  at the scale of $m_t$. 
Based on the renormalization invariance of the product of the bare mass and the condensate,  
we estimated it at the scale of physical $m_t$. 
Our most conservative estimate turned out to be very small $m_{\pi_t} < 70\,\GeV$
which is universal to generic TC2 model.
This would give a serious impact on the phenomenology of the generic TC2 model 
and similar models having two kinds of NG bosons, 
one linear combination of which is absorbed into $W,Z$ bosons and the rest remaining as pseudo NG bosons, 
particularly when they are produced by the NJL-type dynamics.
  
\section*{Acknowledgments}
We would like to thank  Bogdan Dobrescu, Kazumoto Haba, Kyougchul Kong and Shinya Matsuzaki 
for helpful discussions. Special thanks go to Liz Simmons for very helpful comments on the estimation of 
top-pion mass bound. 
This work is supported in part by the JSPS Grant-in-Aid for Scientific 
Research (B) 18340059, The Mitsubishi Foundation, and Daiko Foundation. 
H.F. is supported by the JSPS Research Fellowship for 
Young Scientists and Grant-in-Aid 
for  Scientific Research No.19-99.
K.Y. acknowledges the hospitality and support 
of the Radcliffe Institute for Advanced Study, Harvard University during the completion of the present work.


\begin{thebibliography}{99}
\bibitem{Weinberg:1975gm}
  S.~Weinberg,
  Phys.\ Rev.\ D {\bf 13}, 974 (1976); Phys.\ Rev.\ D {\bf 19}, 1277 (1979);
  L.~Susskind,
  Phys.\ Rev.\ D {\bf 20}, 2619 (1979).
  
\bibitem{Dimopoulos:1979es}
  S.~Dimopoulos and L.~Susskind,
  Nucl.\ Phys.\ B {\bf 155}, 237 (1979);
  E.~Eichten and K.~D.~Lane,
  Phys.\ Lett.\ B {\bf 90}, 125 (1980).
  
\bibitem{Yamawaki:1982tg}
  K.~Yamawaki and T.~Yokota,
  Phys.\ Lett.\  B {\bf 113} (1982), 293;
  Nucl.\ Phys.\  B {\bf 223} (1983), 144. 
 
\bibitem{Peskin:1990zt}
  M.~E.~Peskin and T.~Takeuchi,
  Phys.\ Rev.\ Lett.\  {\bf 65}, 964 (1990);
  B.~Holdom and J.~Terning,
  Phys.\ Lett.\ B {\bf 247}, 88 (1990);
  M.~Golden and L.~Randall,
  Nucl.\ Phys.\ B {\bf 361}, 3 (1991).

\bibitem{Holdom:1984sk}
  B.~Holdom,
  Phys.\ Lett.\ B {\bf 150}, 301 (1985).

\bibitem{Yamawaki:1985zg}
  K.~Yamawaki, M.~Bando and K.~Matumoto,
  Phys.\ Rev.\ Lett.\  {\bf 56}, 1335 (1986);
  T.~Akiba and T.~Yanagida,
  Phys.\ Lett.\ B {\bf 169}, 432 (1986);
  T.~W.~Appelquist, D.~Karabali and L.~C.~R.~Wijewardhana,
  Phys.\ Rev.\ Lett.\  {\bf 57}, 957 (1986).

\bibitem{Miransky:vk}
 V.~A.~Miransky, {\it Dynamical Symmetry Breaking 
 in Quantum Field Theories} (World Scientific Pub. Co., Singapore 1993).

\bibitem{Yamawaki:1996vr}
  K.~Yamawaki,
  in {\it Proceedings of 14th Symposium on Theoretical Physics 
  ``Dynamical Symmetry Breaking and Effective Field Theory''}, 
  Cheju Island, Korea, July 21-26, 1995, ed. J.~E.~Kim 
  (Minumsa Pub. Co., Korea, 1996) p.43-86.

\bibitem{Hill:2002ap}
  C.~T.~Hill, and E.~H.~Simmons,
  Phys. Rept. {\bf 381}, 235 (2003),
  [Erratum {\it ibid.}  {\bf 390}, 553 (2004)].

\bibitem{Appelquist:1996dq}
  T.~Appelquist, J.~Terning and L.~C.~R.~Wijewardhana,
  Phys.\ Rev.\ Lett.\  {\bf 77}, 1214 (1996);
  T.~Appelquist, A.~Ratnaweera, J.~Terning and L.~C.~R.~Wijewardhana,
  Phys.\ Rev.\  D {\bf 58}, 105017 (1998).
  
\bibitem{Miransky:1996pd}
  V.~A.~Miransky and K.~Yamawaki,
  Phys.\ Rev.\  D {\bf 55}, 5051 (1997)
  [Erratum-ibid.\  D {\bf 56}, 3768 (1997)]
  
\bibitem{Banks:1981nn}
  T.~Banks and A.~Zaks,
  Nucl.\ Phys.\  B {\bf 196}, 189 (1982).

\bibitem{Hong:2004td}
  D.~K.~Hong, S.~D.~H.~Hsu and F.~Sannino,
  Phys.\ Lett.\  B {\bf 597}, 89 (2004)
  
\bibitem{Iwasaki:2003de}
  Y.~Iwasaki, K.~Kanaya, S.~Kaya, S.~Sakai and T.~Yoshie,
  Phys.\ Rev.\  D {\bf 69}, 014507 (2004)
  
\bibitem{Appelquist:1998xf}
  T.~Appelquist and F.~Sannino,
  Phys.\ Rev.\ D {\bf 59}, 067702 (1999);

\bibitem{Harada:2005ru}
  M.~Harada, M.~Kurachi and K.~Yamawaki,
  Prog.\ Theor.\ Phys.\  {\bf 115}, 765 (2006)

\bibitem{Hong:2006si}
  D.~K.~Hong and H.~U.~Yee,
  Phys.\ Rev.\ D {\bf 74}, 015011 (2006);
  M.~Piai,
  arXiv:hep-ph/0608241; 
  K.~Haba, S.~Matsuzaki and K.~Yamawaki, in preparation.

\bibitem{Appelquist:1993sg}
  T.~Appelquist and J.~Terning,
  Phys.\ Rev.\  D {\bf 50}, 2116 (1994)
  
\bibitem{Appelquist:2003hn}
  T.~Appelquist, M.~Piai and R.~Shrock,
  Phys.\ Rev.\  D {\bf 69}, 015002 (2004)
  
\bibitem{Chivukula:1988qr}
  R.~S.~Chivukula,
  Phys.\ Rev.\ Lett.\  {\bf 61}, 2657 (1988).

\bibitem{Miransky:1988xi} 
V.~A.~Miransky, M. Tanabashi, and K. Yamawaki,
  Phys. Lett. {\bf B221}, 177 (1989);
  Mod. Phys. Lett. {\bf A4}, 1043 (1989).

\bibitem{Nambu89}
  Y.~Nambu, Enrico Fermi Institute Report No. 89-08, 1989 
  (unpublished); in {\it Proceedings of the 1989 Workshop 
   on Dynamical Symmetry Breaking}, edited by T. Muta and K. Yamawaki 
  (Nagoya University, Nagoya, Japan, 1990).

\bibitem{Marciano89}
  W.~J.~Marciano,
  Phys. Rev. Lett. {\bf 62}, 2793 (1989).

\bibitem{BHL90}
  W.~A.~Bardeen, C.~T.~Hill, and M.~Lindner,
  Phys. Rev. {\bf D41}, 1647 (1990).

\bibitem{Hill:1994hp}
  C.~T.~Hill,
  Phys.\ Lett.\  B {\bf 345}, 483 (1995);

\bibitem{Hill:1991at}
  C.~T.~Hill,
  Phys.\ Lett.\ B {\bf 266}, 419 (1991);
  G.~Buchalla, G.~Burdman, C.~T.~Hill and D.~Kominis,
  Phys.\ Rev.\ D {\bf 53}, 5185 (1996).

\bibitem{Lane:1995gw}
  K.~D.~Lane and E.~Eichten,
  Phys.\ Lett.\  B {\bf 352}, 382 (1995);
  R.~S.~Chivukula, A.~G.~Cohen and E.~H.~Simmons,
  Phys.\ Lett.\  B {\bf 380}, 92 (1996);
  K.~D.~Lane,
  Phys.\ Lett.\  B {\bf 433}, 96 (1998);
  M.~B.~Popovic and E.~H.~Simmons,
  Phys.\ Rev.\  D {\bf 58}, 095007 (1998).

\bibitem{Kikukawa:1989fw}
 Y.~Kikukawa and K.~Yamawaki,
 Phys. Lett. {\bf B234}, 497 (1990).
 
\bibitem{Kondo:1992sq}
 K.-I.~Kondo, M.~Tanabashi, and K.~Yamawaki, 
 Prog. Theor. Phys. {\bf 89}, 1249 (1993),
 Earlier version is  
 K.-I.~Kondo, M.~Tanabashi, and K.~Yamawaki, DPNU-91-20/CHIBA-EP-53 (July, 1991),
 Mod. Phys. Lett. {\bf A8}, 2859 (1993).
 
 \bibitem{Kondo:1991yk}
 K.-I. Kondo, S. Shuto, and K.Yamawaki,  
 Mod. Phys. Lett. {\bf A6}, 3385 (1991);
 N.~V.~Krasnikov,
 Mod. Phys. Lett. {\bf A8}, 797 (1993);
 M.~Harada, Y.~Kikukawa, T.~Kugo, and H.~Nakano,
 Prog. Theor. Phys. {\bf 92}, 1161 (1994).

\bibitem{Dashen:1969eg}
  R.~F.~Dashen,
  Phys.\ Rev.\  {\bf 183}, 1245 (1969).

\bibitem{Pagels:1979hd}
  H.~Pagels and S.~Stokar,
  Phys.\ Rev.\  D {\bf 20}, 2947 (1979).

\bibitem{Bertram:1998wf}
  I.~Bertram and E.~H.~Simmons,
  Phys.\ Lett.\  B {\bf 443}, 347 (1998).

\bibitem{Braam:2007pm}
  F.~Braam, M.~Flossdorf, R.~S.~Chivukula, S.~Di Chiara and E.~H.~Simmons,
  arXiv:0711.1127 [hep-ph].
  
\bibitem{Gusynin:1982kp}
  V.~P.~Gusynin, V.~A.~Miransky and Yu.~A.~Sitenko,
  Phys.\ Lett.\  B {\bf 123}, 407 (1983);
  Y.~Kikukawa and N.~Kitazawa,
  Phys.\ Rev.\  D {\bf 46}, 3117 (1992).
  
\bibitem{Appelquist:1988yc}
  T.~Appelquist, K.~D.~Lane and U.~Mahanta,
  Phys.\ Rev.\ Lett.\  {\bf 61}, 1553 (1988).

\bibitem{Kondo:1988qd}
  K.-I.~Kondo, H.~Mino, and K.~Yamawaki, 
  Phys.\ Rev.\  D {\bf 39}, 2430 (1989);
  K.~Yamawaki, 
  in {\it Proc. Johns Hopkins Workshop on Current Problems in Particle 
  Theory 12, Baltimore, June 8-10, 1988}, 
  edited by G. Domokos and S. Kovesi-Domokos 
  (World Scientific Pub. Co., Singapore 1988);
  T.~Appelquist, M.~Soldate, T.~Takeuchi, and L.~C.~R.~Wijewardhana, 
  {\it ibid}.

\bibitem{Yao:2006px}
  W.~M.~Yao {\it et al.}  [Particle Data Group],
  J.\ Phys.\ G {\bf 33}, 1 (2006).

\bibitem{Nonoyama:1989dq}
  T.~Nonoyama, T.~B.~Suzuki and K.~Yamawaki,
  Prog.\ Theor.\ Phys.\  {\bf 81}, 1238 (1989).

\bibitem{Appelquist:1997fp}
  T.~Appelquist, J.~Terning and L.~C.~R.~Wijewardhana,
  Phys.\ Rev.\ Lett.\  {\bf 79}, 2767 (1997).

\end{thebibliography}
\end{document}